 \documentclass[prd,showpacs,preprintnumbers,amsmath,amssymb,floatfix]{revtex4}

\usepackage{graphicx}
\usepackage{dcolumn}
\usepackage{bm}
\usepackage{amssymb}
\usepackage{epsfig}
\usepackage{color}

\newcommand{\ba}{\begin{eqnarray}}
\newcommand{\ea}{\end{eqnarray}}
\newcommand{\be}{\begin{equation}}
\newcommand{\ee}{\end{equation}}
\newcommand{\bdisplay}{\begin{displaymath}}
\newcommand{\edisplay}{\end{displaymath}}

\newcommand{\eq}[1]{Eq.\,(\ref{#1})}

\newcommand{\calF}{\mbox{${\cal F}$}}
\newcommand{\calG}{\mbox{${\cal G}$}}
\newcommand{\frakg}{\mathfrak{g}}

\newcommand{\alphabold}{\mbox{\small\boldmath $\alpha$}}
\newcommand{\xbold}{\mbox{\boldmath $x$}}

\newcommand{\delchisq}{\Delta \chi^2_i(x_i;\alphabold)}
\newcommand{\delchi}{\Delta \chi^2_i}

\newcommand{\delchimax}{{\delchi}_{\rm max}}

\newcommand{\la}{\,\raisebox{-.8ex}{\,$\stackrel{\textstyle <}{\sim}$}\,\,}

\begin{document}

\title{Analytic derivation of the leading-order gluon distribution function $G(x,Q^2)=xg(x,Q^2)$ from the proton structure function $F_2^{\gamma p}(x,Q^2)$.   Effect of heavy quarks. }

\author{Martin~M.~Block}
\affiliation{Department of Physics and Astronomy, Northwestern University, 
Evanston, IL 60208}
\author{Loyal Durand}
\affiliation{Department of Physics, University of Wisconsin, Madison, WI 53706}
\date{\today}

\begin{abstract}
We extend our previous derivation of  an exact expression for the leading-order (LO) gluon distribution function $G(x,Q^2)=xg(x,Q^2)$ from the DGLAP evolution equation for the proton structure function $F_2^{\gamma p}(x,Q^2)$ for deep inelastic $\gamma^* p$ scattering to include the effects of heavy-quark masses.  We derive the equation for $G(x,Q^2)$ in two different ways, first using our original differential-equation method, and then using a new method based on Laplace transforms. The results do not require the use of the gluon evolution equation, or, to good approximation,  knowledge of the individual quark distributions. Given an analytic expression that successfully reproduces the known  experimental data for $F_2^{\gamma p}(x,Q^2)$ in a domain ${\cal D}(x,Q^2)$---where $x_{\rm min}(Q^2) \le x \le x_{\rm max}(Q^2)$, $Q^2_{\rm min}\le Q^2\le  Q^2_{\rm max}$ of the Bjorken variable $x$  and the virtuality $Q^2$---$G(x,Q^2)$ is uniquely determined in the same domain. 
As an application of the method, we construct a new global parametrization of the complete set of ZEUS data on $F_2^{\gamma p}(x,Q^2)$, and use this to determine the 5 quark gluon distribution,  $G(x,Q^2)$,  for massless $u,\,d,\,s$ and massive $c,\,b$ quarks and discuss the mass effects evident in the result. We compare these results to the gluon distributions for CTEQ6L, and in the domain ${\cal D}(x,Q^2)$ where they should agree, they do not; the discrepancy is due to the fact that  the CTEQ6L results do not give an accurate description of the ZEUS  $F_2^{\gamma p}(x,Q^2)$ experimental data. We emphasize that our method for obtaining the the LO gluon distribution  connects $G(x,Q^2)$ {\em directly} to the proton structure function without  either the need for individual parton distributions or the gluon evolution equation. 
\end{abstract}


\maketitle


\section{Introduction} \label{sec:introduction} 

The quark and gluon distributions in hadrons  play a key role in our understanding of Standard Model processes, in our predictions for such processes at accelerators, and in our searches for new physics.  In particular, accurate knowledge of gluon distribution functions at small Bjorken $x$ will play a vital role in estimating backgrounds, and hence, our ability to search for new physics at the Large Hadron Collider.  

The gluon and quark distribution functions have traditionally been determined  simultaneously by fitting experimental data  (mainly at small $x$) on the  proton structure function $F_2^{\gamma p}(x,Q^2)$ measured in deep inelastic $ep$ (or $\gamma^*p$) scattering, over a large domain  of values of $x$ and $Q^2$.  The process starts with an initial  $Q^2_0$, typically in the 1 to 2 GeV$^2$ range, and  individual quark and gluon trial distributions given as functions of $x$. The distributions are evolved to larger $Q^2$ using the coupled integral-differential DGLAP equations  \cite{dglap1,dglap2,dglap3}, and the results used to predict the measured quantities. The final distributions are then determined by adjusting the parameters in the input to obtain a best fit to the data. For recent determinations of the gluon and quark distributions, see \cite{CTEQ6.1,CTEQ6.5,MRST,MRST4}.  

This procedure is rather indirect, especially so in the case of the gluon: the gluon distribution $G(x,Q^2)$ does not appear in the experimentally accessible quantity $F_2^{\gamma p}(x,Q^2)$, and is determined only through the quark distributions in conjunction with the evolution equations.  It is  further not clear without detailed analysis \cite{MRST3,CTEQchi2,CTEQ_Hessian,CTEQ6.5} how sensitive the results are to the parametrizations of the initial parton distributions, or how well the gluon distribution is actually determined.

In two recent papers \cite{bdm1,bdm2} papers, we presented a new method for determining the gluon distribution function $G(x,Q^2)=xg(x,Q^2)$ in leading order (LO) in QCD directly from a global parametrization of the data on  $F_2^{\gamma p}(x,Q^2)$, and applied the results to obtain sensitive tests of some existing parton distributions. The method uses only the LO DGLAP evolution equation \cite{dglap1,dglap2,dglap3} for $F_2^{\gamma p}(x,Q^2)$ in the usual approximation in which the active quarks are treated as massless. In contrast to previous methods for determining $G(x,Q^2)$, it  does not require knowledge of the separate quark distributions in the region in which structure function data exist, nor does it require the use of the evolution equation for $G(x,Q^2)$, both considerable simplifications. 

We generalize here to the case of massive quarks using the same methods and the less-stringent approximation that mass effects are neglected only in terms found  to be small.  The additional complications are then minimal. Furthermore, to good approximation, detailed knowledge of the individual quark distributions is still not required. 

In the following Sections, we present two analytic methods that determine $G(x,Q^2)$ directly from $F_2^{\gamma p}(x,Q^2)$. The first involves an inhomogeneous second-order linear differential equation for $G(x,Q^2)$ derived from the LO DGLAP equation for the evolution of $F_2^{\gamma p}(x,Q^2)$ as a function of $Q^2$.  The inhomogeneous driving term in the equation is determined entirely by $F_2^{\gamma p}(x,Q^2)$ and its derivatives, and the equation can be solved explicitly. The second method involves a somewhat unusual application of Laplace transforms to the DGLAP equation, and leads to equivalent results without the intervening differential equation.
 
Our method, in either form, determines the LO gluon distribution function $G(x,Q^2)$ in terms of  the measured structure function $F_2^{\gamma p}(x,Q^2)$.  The result is unique, within experimental uncertainties, in the domain ${\cal D}(x,Q^2)$  in which there are experimental data and requires no  assumptions about the initial shapes of the  gluon or the individual quark distributions. Furthermore, it is not necessary to use the DGLAP evolution equation for $G(x,Q^2)$ as long as one remains in the experimental region, though it provides a useful consistency check.  These LO  results are, of course,  uncertain at the level of the next-to-leading (NLO ) contributions to the DGLAP equations; these corrections will be considered elsewhere.

As an application of the method, we construct an accurate new global parametrization of the complete set of ZEUS data on $F_2^{\gamma p}(x,Q^2)$ \cite{ZEUS1,ZEUS2}, and use this to determine the 5-quark gluon distribution, $G(x,Q^2)$, for massless $u,\,d,\,s$ and massive $c,\,b$ quarks.  We then  discuss in detail the mass effects evident in a comparison with the results for massless quarks, including their interesting dependence on $Q^2$. 

We note that the same approach can be used to relate $G(x,Q^2)$ directly to data on the remaining DIS structure functions  $F_{2(3)}^{\gamma}$, $F_{2(3)}^{\gamma Z}$, $F_{2(3)}^{Z}$, and$F_{2(3)}^{W^\pm}$ for neutral- and charged-current scattering,. 

It is still necessary to obtain the individual quark distributions in order to predict some quantities, or to check the accuracy of our approximations.  As noted above, this is traditionally done using simultaneous fits to all quark and gluon distributions and the complete set of coupled DGLAP equations, fits which may use jet or other quark- or gluon-dependent data. However, our results show very clearly the direct connection of $G(x,Q^2)$ to the accurate deep inelastic scattering data and should provide a useful check on the gluon distributions obtained using other methods. We show, for example, that the disagreement of our LO  results for massless $u,\,d,\,s$ and $c$ quarks with the gluon distribution obtained in the CTEQ6L fits \cite{CTEQ6L}  results from the failure of the latter to reproduce accurately the ZEUS results for $F_2^{\gamma p}(x,Q^2)$.


\section{Analytic treatment of the leading order gluon distribution function}\label{sec:diffeq}

\subsection{Treatment of heavy-quark effects}\label{subsec:heavy_quarks} 

In the presence of heavy quarks, here taken as $c$ and $b$ quarks, the DGLAP evolution equations for the quark distribution functions $q_i(x,Q^2)$, $i\in u,\bar{u},d,\bar{d}, s,\bar{s},c,\bar{c},b,\bar{b}$, must be modified to take into account the effect of production thresholds for pairs of heavy quarks, assuming none are present in the initial hadron, and  mass effects in the parton splitting functions\,; each affects the $Q^2$ evolution of the parton distributions. We will treat these effects using a simplified version of the method introduced by Aivazis, Collins, Olness, and Tung (ACOT) \cite{ACOT}.  In this method 
\cite{AOT,ACOT,S-ACOT,Tung_masses,CTEQ6.5},  a heavy quark  $q_i$  has its  usual Bjorken scaling variable $x$ replaced in its splitting functions $K_{gq_i}(x/z)$ and $K_{q_iq_i}(x/z)$   by the ``slow'' scaling variable $x_i=x(1+4M_i^2/Q^2)$ that appears naturally. Further mass effects in the splitting functions are ignored, so those functions retain the forms used for massless quarks. The integrations in the evolution equations are then taken to run from $z=x_i$ to $z=1$ rather than $z=x$ to $z=1$, thus imposing the threshold condition $x<x_i\leq 1$. (There are some additional changes if one is interested specifically in the production of heavy quarks near threshold \cite{Tung_masses,CTEQ6.5}, but these are not important for inclusive deep inelastic scattering and will be ignored here.)

With the modifications above, the evolution equation for a heavy quark becomes
\be
\label{q_evolution}
\frac{\partial q_i(x,Q^2)}{\partial\ln Q^2} = \frac{\alpha_s}{4\pi}\int_{x_i}^1\frac{dz}{z}\left[q_i(z,Q^2)K_{qq}\left(\frac{x_i}{z}\right)+g(z,Q^2)K_{gq}\left(\frac{x_i}{z}\right)\right].
\ee
Multiplying by the squares of the quark charges $e_i^2$ and summing over the quarks $i=u,\bar{u},d,\bar{d}, s,\ldots$ to get $F_2^{\gamma p}(x,Q^2)/x=\sum_ie_i^2q_i(x,Q^2)$ on the left, we find that
\be
\label{F2_evolution}
\frac{1}{x}\frac{\partial F_2^{\gamma p}(x,Q^2)}{\partial\ln Q^2} = \frac{\alpha_s}{4\pi}\left[\sum_i\int_{x_i}^1\frac{dz}{z}e_i^2q_i(z,Q^2)K_{qq}\left(\frac{x_i}{z}\right) +\sum_ie_i^2 \int_{x_i}^1\frac{dz}{z^2}G(z,Q^2)K_{gq}\left(\frac{x_i}{z}\right)\right].
\ee

All terms in the second sum on the right involve $G(z,Q^2) = zg(z,Q^2)$, with only the arguments in the splitting functions and the ranges of integration modified in the heavy quark terms.  This will cause no problems as will be seen below.

The heavy-quark terms in the first sum on the right do cause a problem: because of the different arguments in the the splitting functions and the different ranges of integration, the sum is not expressible in terms of $F_2^{\gamma p}$ as would be the case for massless quarks. This becomes clearer if we introduce the parameters $\eta_i(Q^2)=1+(4M_i^2/Q^2)$ and change the variable of integration in each term from $z$ to $z'=z/\eta_i$, and then drop the prime for simplicity. The lower limit of the integral becomes  $x$ and the upper limit can be extended  from $1/{\eta_i}$ to $1$, since $q_i(z)\equiv 0$ for $z\ge 1.$ The quark term then becomes
\be
\label{quark_sum}
\frac{\alpha_s}{4\pi}\int_x^1\frac{dz}{z}\sum_i e_i^2 q_i\left(\eta_i z,Q^2\right) K_{qq}\left(\frac{x}{z}\right) \equiv \frac{\alpha_s}{4\pi}\int_x^1\frac{dz}{z^2}F_{2,\rm shifted}^{\gamma p}(z,Q^2)K_{qq}\left(\frac{x}{z}\right), 
\ee
where $K_{qq}$ is the common splitting function for massless quarks and $F_{2,\rm shifted}^{\gamma p}(z,Q^2) = \sum_i e_i^2 zq_i(\eta_i z,Q^2)$. When the quarks are taken as massless, $\eta_i=1$, and the quark sum is just $F_2^{\gamma p}(z,Q^2)/z$. This is no longer true when heavy quarks are present, and information on the individual quark distributions is clearly needed to evaluate the sum exactly. 

The methods developed below for determining $G(x,Q^2)$ exactly from Eq.\ (\ref{F2_evolution}) require only that $F_z^{\gamma p}$ and $F_{2, \rm shifted}^{\gamma p}$ be known. This is of course the case when the quark distributions are known.  If they are not, we can still proceed using reasonable approximations.
The simplest is to replace $\eta_i$ by 1 in Eq.\ (\ref{quark_sum}), but not in the gluon terms in Eq.\ (\ref{F2_evolution}) where threshold effects are important.  The quark sum in Eq.\ (\ref{quark_sum}) then becomes $F_2^{\gamma p}(z,Q^2)$, and we obtain the simplified evolution equation
\be
\label{F2_evolution2}
\frac{1}{x}\frac{\partial F_2^{\gamma p}(x,Q^2)}{\partial\ln Q^2}  = \frac{\alpha_s}{4\pi}\left[\int_{x}^1\frac{dz}{z^2}F_2^{\gamma p}(z,Q^2)K_{qq}\left(\frac{x}{z}\right) +\sum_ie_i^2\frac{1}{\eta_i} \int_{x}^1\frac{dz}{z^2}G(\eta_iz,Q^2)K_{gq}\left(\frac{x}{z}\right)\right].
\ee
This approximation should be good.  The factor $F_2^{\gamma p}(z,Q^2)$ in the first integral on the right will be evaluated using the experimentally determined structure function so automatically includes the threshold conditions for the activation of heavy quarks and the expected suppression of the heavy quark distributions. The actual approximation is  the neglect of the shift from $z$ to $\eta_iz$ for the point at which $q_i$ is evaluated.   This should not be a large effect for $Q^2$ sufficiently large and $q_i(x,Q^2)$ not too rapidly varying in $x$.  In addition, the contribution of the $F_2^{\gamma p}$-dependent terms in Eq.\ (\ref{F2_evolution2}) to our final expression for $G(x,Q^2)$ is considerably smaller at small $x$ than that dependent on $\partial F_2^{\gamma p}(x,Q^2)/\partial\ln Q^2$, significantly  reducing the importance of the approximation.

We have checked the validity of the approximation using the CTEQ6.5 \cite{CTEQ6.5}  quark distributions which were derived using the simplified ACOT method to treat mass effects as is done here.  The errors introduced by replacing the proper quark sum in Eq. (\ref{quark_sum})  by $F_2^{\gamma p}(x,Q^2)$
is less than 14\%, 9\%, and 2.5\% of $F_2^{\gamma p}$ for $Q^2= 5, \,20$ and 100 GeV$^2$, respectively, for $0.1>x>10^{-4}$. Taking into account the expected suppression of the $F_2^{\gamma p}$-dependent terms in the final result relative to those dependent on  $\partial F_2^{\gamma p}(x,Q^2)/\partial\ln Q^2$, we expect associated errors in $G(x,Q^2)$ considerably smaller. They are in fact less than 2--3\% in the final  $G$.

A  more accurate approximation for the input that attempts to incorporate the shift from $z$ to $\eta_c z$ and $\eta_b z$ in the $c$- and $b$-quark distribution functions in Eq.\ (\ref{quark_sum}) is to equate those terms to their maxima of 2/3 and 1/6 of the light-quark sea contribution to $F_2^{\gamma p}$, then shift the arguments appropriately. In particular, taking into account the shifts of the heavy-quark variables in Eq.\ (\ref{q_evolution}), we approximate the $c$ and $b$ terms in $F_2^{\gamma p}$ as
\ba
\label{c_approx1}
xc(x,Q^2) &\approx&  \frac{3}{4\eta_c}F_{\rm light}^{\gamma p}(\eta_c x,Q^2), \\
\label{b_approx1}
x b(x,Q^2) &\approx& \frac{3}{4\eta_b}F_{\rm light}^{\gamma p}(\eta_b x,Q^2).
\ea
Then, noting that the total sea distribution is $F_{\rm sea}^{\gamma p}=F_{\rm light}^{\gamma p}+(8/9)xc+(2/9)xb$, using the approximations above, and solving for $F_{\rm light}^{\gamma p}$, we find that 
\ba
\label{c_approx2}
xc(x,Q^2) &\approx& \frac{3}{4\eta_c}\left[1+\frac{2}{3\eta_c}T(\eta_c)+\frac{1}{6\eta_b}T(\eta_b)\right]^{-1}F_{\rm sea}^{\gamma p}(\eta_c x,Q^2), \\
\label{b_approx2}
xb(x,Q^2) &\approx& \frac{3}{4\eta_b}\left[1+\frac{2}{3\eta_c}T(\eta_c)+\frac{1}{6\eta_b}T(\eta_b)\right]^{-1}F_{\rm sea}^{\gamma p}(\eta_b x,Q^2).
\ea
Here $T(\eta)$ is a scaling operator with the property that $T(\eta)f(x)=f(\eta x)$.  As will be seen in detail later, $T$ can also be viewed as a  translation operator in the variable $v=\ln(1/x)$ that shifts $v$ to $v-\ln\eta$, and can be approximated for $Q^2$ not too small by the unit operator.

The problem at this point is that $F_{\rm sea}^{\gamma p}$ is not known directly from data.  We can estimate it by starting with $F_2^{\gamma p}$ at small $x$, where the valence-quark contributions are small relative to the sea, and making a smooth extrapolation to zero for $x\rightarrow 1$ using a reasonable model for the valence distribution.  More simply, we can just replace $F_{\rm sea}^{\gamma p}$ by $F_2^{\gamma p}$, a good approximation at small $x$ and large $Q^2$. In either case, we can use the approximations in Eqs.\ (\ref{c_approx2}) and (\ref{b_approx2}) to estimate the effect of the shifts from $z$ to $\eta_c z$ and $\eta_b z$ in Eq.\ (\ref{F2_evolution}). If we use $F_2^{\gamma p}$, we find that
\ba
F_{2,\rm shifted}^{\gamma p}(x,Q^2) &\equiv& \sum_i e_i^2 xq_i(\eta_i x,Q^2)
\approx  F_2^{\gamma p}(x,Q^2) + \left[1+\frac{2}{3\eta_c}T(\eta_c)+\frac{1}{6\eta_b}T(\eta_b)\right]^{-1} \nonumber \\
\label{F2_shiftapprox}
&& \times\left[\frac{2}{3\eta_c}\left(\frac{1}{\eta_c}F_2^{\gamma p}(\eta_c^2 x,Q^2)-F_2^{\gamma p}(\eta_c x,Q^2)\right) + \frac{1}{6\eta_c}\left(\frac{1}{\eta_b}F_2^{\gamma p}(\eta_b^2 x,Q^2)-F_2^{\gamma p}(\eta_b x,Q^2)\right)\right].
\ea
This approximation is quite good at small $x$.

Because of the additional complication in intermediate results, we will use the simpler approximation for the evolution equation in Eq.\ (\ref{F2_evolution2}), and discuss the accuracy of the result later.


\subsection{ Differential equation for $G(x,Q^2)$}\label{subsec:diffeq_derived}

To obtain a differential equation for $G(x,Q^2)$, we follow the procedure used in our earlier work \cite{bdm1}, but with the modified $G$-dependent term in Eq.\ (\ref{F2_evolution2}) instead of that for massless quarks. Writing out the sum and the splitting function explicitly for massless $u,\,d,\ \textrm{and}\, s$ quarks and massive $c$ and $b$ quarks, we obtain the equation
\ba
\frac{4}{3}\int_x^1\frac{dz}{z^2}G(z,Q^2)\left(\frac{x^2+(z-x)^2}{z^2}\right) + \frac{8}{9}\int_{x_c}^1 \frac{dz}{z^2}G(z,Q^2)\left(\frac{x_c^2+(z-x_c)^2}{z^2}\right) \hspace*{0.5in} && \nonumber \\
\label{G_integral}
+ \frac{2}{9}\int_{x_b}^1 \frac{dz}{z^2}G(z,Q^2)\left(\frac{x_b^2+(z-x_b)^2}{z^2}\right)=\frac{4\pi}{\alpha_s}\frac{1}{x}{\cal F}_2^{\gamma p}(x,Q^2).&&
\ea
Here  $x_i=x\eta_i$, $\eta_i=1+(4M_i^2/Q^2)$, and ${\cal F}_2^{\gamma p}(x,Q^2)$ is the sum of the $F_2^{\gamma p}$-dependent terms in Eq.\ (\ref{F2_evolution2}),
\begin{subequations}
\ba
\label{calF2_1}
{\cal F}_2^{\gamma p}(x,Q^2) &=&\frac{\partial F_2^{\gamma p}(x,Q^2)}{\partial \ln (Q^2)}-\frac{\alpha_s}{4\pi}\left\{ 4{F_2^{\gamma p}(x,Q^2)}\!+\!\frac{16}{3}\left[{F_2^{\gamma p}(x,Q^2)}\ln\frac{1-x}{x}\right.\right.\nonumber\\
&&\left.\left.
+x\int_x^1\left(\frac{F_2^{\gamma p}(z,Q^2)}{z}-\frac{F_2^{\gamma p}(x,Q^2)}{x}\right)\frac{dz}{ z-x}\right]-\frac{8}{3}x\int_x^1F_2^{\gamma p}(z,Q^2)\left(1+\frac{x}{z}\right)\frac{\,dz}{z^2}\right\} \\
&=& \frac{\partial{\cal F}_2^{\gamma p}(x,Q^2)}{\partial\ln Q^2} 
 -\frac{\alpha_s}{4\pi}\left\{\frac{16}{3}\int_x^1\frac{\partial F_2^{\gamma p}}{\partial z}(z,Q^2)\,\ln\frac{z}{z-x}dz \right. \nonumber \\
&& \left.  -\frac{4}{3}\int_x^1\frac{\partial F_2^{\gamma p}}{\partial z}(z,Q^2)\left(\frac{x^2}{z^2}+\frac{2x}{z}\right)dz\right\}. \label{calF2}
\ea
\end{subequations}
The only change if we  treat the quark sum in Eq.\ (\ref{quark_sum}) exactly, or in an approximation such as that in Eq.\ (\ref{F2_shiftapprox}), is the replacement of $F_2^{\gamma p}$ by $F_{2,\rm shifted}^{\gamma p}$ in all but the first term on the right-hand side of this equation. We will therefore assume that ${\cal F}_2^{\gamma p}$ is known.

The three terms on the left hand side of Eq.\ (\ref{G_integral}) can be combined with a common splitting function by rescaling the variables $z$ in the second and third terms to $\eta_c z$ and $\eta_b z$, respectively. This gives the modified equation 
\be
\label{G_integral2}
\int_x^1\frac{dz}{z^2}\left(\frac{4}{3}G(z,Q^2) + \frac{8}{9}\frac{1}{\eta_c}G(\eta_c z,Q^2)+ \frac{2}{9}\frac{1}{\eta_b}G(\eta_b z,Q^2)\right)\left(\frac{x^2+(z-x)^2}{z^2}\right)=\frac{4\pi}{\alpha_s}\frac{1}{x}{\cal F}_2^{\gamma p}(x,Q^2).
\ee

 As was observed in \cite{bdm1}, a three-fold differentiation of Eq.\ (\ref{G_integral2}) with respect to $x$ eliminates the integration on the left hand side.  Dividing out the factor 4/3 in the first term from the sum of the light-quark charges in Eq.\ (\ref{G_integral}) and multiplying by $x^4$, we obtain an inhomogeneous second-order differential equation for $G(x,Q^2)$,
\be
\label{scaled_diff_eq}
\left(x^2\frac{\partial^2}{\partial x^2}-2x\frac{\partial}{\partial x}+4\right)\left(G(x,Q^2)+\frac{2}{3}\frac{1}{\eta_c} G(x_c,Q^2)+\frac{1}{6}\frac{1}{\eta_b} G(x_b,Q^2)\right) = {\cal G}_3(x,Q^2),
\ee
where
\be
\label{calG_3}
{\cal G}_3(x,Q^2) = -\frac{3}{4}\frac{4\pi}{\alpha_s(Q^2)}x^4\frac{\partial^3({\cal F}_2^{\gamma p}(x,Q^2)/x)}{\partial x^3}.
\ee
The functions $G(x_i,Q^2)$ in Eq.\ (\ref{scaled_diff_eq}) vanish identically for $x_i>1$, conditions that can  be made explicit if so desired by introducing step functions $\theta(1-x_i)$ with $G(x_i,Q^2)\rightarrow\theta(1-x_i)G(x_i,Q^2)$. In order to simplify the appearance of our equations, we will generally suppress these step functions.

The function ${\cal G}_3$ on the right hand side of Eq.\ (\ref{calG_3}) is defined here using the multiplicative factor 3/4 for three massless quarks rather than the factor 9/20 used in our treatment of the problem for the 4 massless $u,\,d,\,s$ and $c$ quarks in Ref. \cite{bdm1}.  More generally, for $n$ effectively massless quarks, we define ${\cal G}_n$ with the factor 3/4 replaced by $1/\sum_i e_i^2$ where the sum runs over these ``light'' quarks and antiquarks. Thus, the function used in \cite{bdm1} for 4 quarks was ${\cal G}_4(x,Q^2) = (3/5){\cal G}_3(x,Q^2)$.

The differential equation in Eq.\ (\ref{scaled_diff_eq}) can be put in simpler form by changing to the new variable $v=\ln(1/x)$. Then with $\hat{G}(v,Q^2)=G(e^{-v},Q^2)$ and $\hat{\cal G}_3(v,Q^2)={\cal G}_3(e^{-v},Q^2)$, 
\be
\label{v_diff_eq}
\left(\frac{\partial^2}{\partial v^2} + 3\frac{\partial}{\partial v} + 4\right)\left(\hat{G}(v,Q^2)+\frac{2}{3}\frac{1}{\eta_c}\hat{G}(v-\ln\eta_c)+\frac{1}{6}\frac{1}{\eta_b}\hat{G}(v-\ln\eta_b)\right)=\hat{\cal G}_3(v,Q^2).
\ee
Solving for  the $v$-dependent function on the left hand side using standard methods,  we find that  the exact solution for the boundary conditions $\hat{G}(0,Q^2)=(\partial\hat{G}/\partial v)(0,Q^2)=0$  is 
\be
\label{v_solution}
\hat{G}(v,Q^2)+\frac{2}{3}\frac{1}{\eta_c}\hat{G}(v-\ln\eta_c)+\frac{1}{6}\frac{1}{\eta_b}\hat{G}(v-\ln\eta_b)=  \hat{\cal H}_3(v,Q^2),
\ee
where $ \hat{\cal H}_3(v,Q^2)$ is the function
\be
\label{Hhat_defined}
 \hat{\cal H}_3(v,Q^2) = \frac{1}{\lambda_+-\lambda_-} \int_0^vdv'\left[e^{\lambda_+(v-v')}-e^{\lambda_-(v-v')}\right]\hat{\calG}_3(v',Q^2). 
\ee
 Here  $\lambda_\pm = k\pm i\omega$ are the roots of the factored form of the differential operator in Eq.\ (\ref{v_diff_eq}):  $D^2+3D+4=(D-\lambda_\pm)(D-\lambda_\mp)$, $D=\partial/\partial v$, with $k=-3/2$ and $\omega=\sqrt{7}/2$. 

Finally, converting back to $x$ as the variable, we find 
\be
\label{x_solution}
G(x,Q^2)+\frac{2}{3}\frac{1}{\eta_c} G(x_c,Q^2)+\frac{1}{6}\frac{1}{\eta_b} G(x_b,Q^2) = {\cal H}_3(x,Q^2)
\ee
where
\begin{subequations}
\ba
\label{H3}
{\cal H}_3(x,Q^2)&=&  \frac{1}{\lambda_+-\lambda_-}\int_x^1  \frac{dz}{z}\left[\left(\frac{z}{x}\right)^{\lambda_+} - \left(\frac{z}{x}\right)^{\lambda_-}\right]{\cal G}_3(z,Q^2)  \\
\label{H3_alt}
&=&  \frac{1}{\omega}\int_x^1\frac{dz}{z}\left(\frac{z}{x}\right)^k\sin\left(\omega\ln\frac{z}{x}\right) {\cal G}_3(z,Q^2) \label{G3int2}.
\ea
\end{subequations}

Two steps remain for us to find a solution for $G(x,Q^2)$, first, the evaluation of the integral of ${\cal G}_3(z,Q^2)$ which gives  ${\cal H}_3(x,Q^2)$, and second, the solution of the resulting mixed expression in $x$, $x_c$, and $x_b$ for $G(x,Q^2)$.


\subsection{Calculation of ${\cal H}_3$}\label{subsec:Hcalc}

After some manipulation starting with the expression for ${\cal F}_2^{\gamma p}$ in Eq.\ (\ref{calF2}), we find that  ${\cal G}_3(x,Q^2)$ can be written as
\ba
{\cal G}_3(x,Q^2) &=& -\frac{3}{4}\frac{4\pi}{\alpha_s}x^4\frac{\partial^3}{\partial x^3}\left(\frac{1}{x}\frac{\partial F_2^{\gamma p}(x,Q^2)}{\partial\ln Q^2}\right) +4x^4\frac{\partial^3}{\partial x^3}\left(\frac{1}{x}\int_x^1dz \frac{\partial F_2^{\gamma p}}{\partial z}(z,Q^2)\ln\frac{z}{z-x}\right) \nonumber \\
&& + \left(4x\frac{\partial}{\partial x}-5x^2\frac{\partial^2}{\partial x^2}+3x^3\frac{\partial^3}{\partial x^3}\right)F_2^{\gamma p}(x,Q^2).\label{calG3}
\ea

Because ${\cal F}_2^{\gamma p}$ is determined by experimental data with limited accuracy, the appearance of high derivatives (up to the fourth) in the expression above could be regarded with suspicion.  However, this apparent problem  largely disappears when we substitute ${\cal F}_2^{\gamma p}$ into  Eq.\ (\ref{H3}). We can then integrate by parts to eliminate these derivatives as much as possible.  The boundary conditions $F_2^{\gamma p}(1,Q^2)=\partial F_2^{\gamma p}(1,Q^2)/\partial x =0$ at $x=1$ eliminate the leading endpoint terms in the integrations.  The use of the expression in Eq.\ (\ref{H3}) rather than that in Eq.\ (\ref{H3_alt}) also makes it easy to use the condition that determines the roots in the factored form of the differential operator in Eq.\ (\ref{v_diff_eq}), $\lambda_\pm^2+3\lambda_\pm+4=0$, to eliminate high powers of $\lambda_\pm$ and reduce the results to simpler form.

Following this procedure, and using the identity for the derivatives of the logarithmic terms given in Eq.\ (11) of Ref. \cite{bdm1}, we find that
\ba
{\cal H}_3(x,Q^2) &=&
  \frac{9}{4}\frac{4\pi}{\alpha_s}\frac{\partial F_2^{\gamma p}}{\partial \ln Q^2}(x,Q^2)-\frac{3}{4}\frac{4\pi}{\alpha_s}x\frac{\partial^2 F_2^{\gamma p}}{\partial x \partial \ln Q^2}(x,Q^2)  \nonumber \\
  &&  -\frac{3}{2}\int_x^1\frac{dz}{z}\left(\frac{z}{x}\right)^k\left[\frac{k+3}{\omega}\sin\omega\ln\frac{z}{x}+\cos\omega\ln\frac{z}{x}\right]\frac{4\pi}{\alpha_s}\frac{\partial F_2^{\gamma p}}{\partial\ln Q^2}(z,Q^2) \nonumber \\
 &&-5F_2^{\gamma p}(x,Q^2)+3x\frac{\partial}{\partial x}F_2^{\gamma p}(x,Q^2) \nonumber \\
 && -12\int_x^1 \frac{\partial F_2^{\gamma p}}{\partial z}(z,Q^2)\ln\frac{z}{z-x}dz \nonumber \\
\label{Fbar_reduction}
&& +4x\int_x^1\frac{\partial^2F_2^{\gamma p}}{\partial z^2}(z,Q^2)\ln\frac{1}{z-x}dz+4x\ln\frac{1}{x}\,\frac{\partial F_2^{\gamma p}}{\partial x}(x,Q^2) \nonumber \\
 &&+\int_x^1\frac{dz}{z}\left(\frac{z}{x}\right)^k\left[\frac{20+12k}{\omega}\sin\omega\ln\frac{z}{x}+12\cos\omega\ln\frac{z}{x}\right]F_2^{\gamma p}(z,Q^2) \nonumber \\
&&
+8\int_x^1\frac{dz}{z}\left(\frac{z}{x}\right)^k\left[\frac{k+3}{\omega}\sin\omega\ln\frac{z}{x}+\cos\omega\ln\frac{z}{x}\right]\int_z^1\frac{\partial F_2^{\gamma p}}{\partial z'}(z',Q^2)\ln\frac{z'}{z'-z}dz'.
 \label{calH_3}
 \ea
While somewhat lengthy, this exact expression has the advantage that it determines the gluon distribution directly in terms of $F_2^{\gamma p}$ and its first two derivatives. Only one double integral remains, a point that simplifies numerical work considerably. This would not be the case if we had simply integrated out the explicit derivatives $(\partial/\partial x)^3$ that act on ${\cal F}_2^{\gamma p}/x$ in Eq.\ (\ref{calG_3}) and had then used the expression for ${\cal F}_2^{\gamma p}$ in Eq.\ (\ref{calF2_1}).

The largest terms in ${\cal H}_3$ at small $x$ are those in the first two lines.  The leading terms in Eq.\ (\ref{calH_3}) are local and can be evaluated directly at the $x$ and $Q^2$ of interest using the global parametrization of $F_2^{\gamma p}(x,Q^2)$. The remaining terms involve at most second derivatives.  These derivatives are very well determined by the global parametrization of $F_2^{\gamma p}(x,Q^2)$. The global nature of the fit is essential here: the derivatives would be considerably less certain if we attempted to determine them using only local data. 
 
The problem now is to evaluate the integrals in Eq.\ (\ref{calH_3}). These include the region $x_P<x<1$, $x_P\approx 0.09$, not covered in the global fit to the $F_2^{\gamma p}$ data of Berger, Block, and Tan \cite{bbt2}, updated later in Sec.\ \ref{subsec:ZEUS} of this paper. In the absence of a global fit that includes data for $x_P<x<1$, the integrals can still be evaluated approximately by using a reasonable extension of $F_2^{\gamma p}$ over that region which satisfies the boundary conditions for $x\rightarrow 1$, matches smoothly to the existing fit for $x<x_P$, and is consistent with the existing data. The results in the small-$x$ region $x\ll x_P$ turn out to be insensitive to the form of the extension.

${\cal H}_3(x,Q^2)$ can be converted to the $v$ form $\hat{\cal H}_3(v,Q^2)$ by the substitution $x=e^{-v}$, $v=\ln(1/x)$ either before or after the integrals are evaluated.


\subsection{Laplace transform method}\label{subsec:Laplace}

Before considering the solution of Eq.\ (\ref{v_solution}), we will rederive  it directly using a Laplace transform method given in \cite{bdm2} which avoids the intermediate steps involved with the differential equation approach, and again shows that high derivatives of the structure functions do not appear in $\hat{\cal H}_3(v,Q^2)$.

We will start with Eq.\ (\ref{G_integral2}). Dividing out the sum of the squares of the light quark charges in the first term and multiplying the equation by $x$, we write this as
\be
\label{G_integral3}
\int_x^1\frac{dz}{z}\frakg(z,Q^2) \frac{x}{z}\left(\frac{x^2+(z-x)^2}{z^2}\right)=\frac{3}{4}\frac{4\pi}{\alpha_s}{\cal F}_2^{\gamma p}(x,Q^2).
\ee
where 
\be
\label{G_tilde}
\frakg(z,Q^2) = G(z,Q^2) + \frac{2}{3}\frac{1}{\eta_c}G(\eta_c z,Q^2)+ \frac{1}{6}\frac{1}{\eta_b}G(\eta_b z,Q^2).
\ee
Changing variables from $x$ and $z$ to $v=\ln(1/x)$ and $w=\ln(1/z)$, Eq.\ (\ref{G_integral3}) and \eq{G_tilde} can be rewritten as
\be
\label{G_tilde_v}
\int_0^v\hat{\frakg}(w,Q^2)\hat{h}(v-w)dw = \hat{f}(v,Q^2),
\ee
where we have introduced the notation $\hat{F}(v,Q^2)$ for the $v$-space form of any function $F(x,Q^2)$, 
\be
\label{hat_form}
\hat{F}(v,Q^2)=F(e^{-v},Q^2).
\ee
The functions in Eq.\  (\ref{G_tilde_v}) are then
\begin{subequations}
\ba
\label{h_defined}
&& \hat{h}(v) = e^{-v}\left(1-2e^{-v}+2e^{-2v}\right), \\
\label{fracg}
&& \hat{\frakg}(v,Q^2)=\hat{G}(v,Q^2) + \frac{2}{3}\frac{1}{\eta_c}\hat{G}(v-\ln\eta_c,Q^2)+ \frac{1}{6}\frac{1}{\eta_b}\hat{G}(v-\ln\eta_b ,Q^2), \\
\label{f_defined}
&& \hat{f}(v,Q^2) =  \frac{3}{4}\frac{4\pi}{\alpha_s}{\cal F}_2^{\gamma p}\left(e^{-v},Q^2\right).
\ea
\end{subequations}

Equation (\ref{G_tilde_v}) relates $\hat{f}(v,Q^2)$, a quantity determined by experiment, to the convolution of $\hat{\frakg}$, a function dependent on the desired function $\hat{G}(v,Q^2)$, with $h(v)$, which is simply $e^{-v}$ times the gluon LO splitting function  for the production of a quark from a gluon, expressed in $v$-space.

The convolution theorem for Laplace transforms applied to transformable  functions $p$ and $q$ states that
\begin{subequations}
\ba
\label{Laplace_direct}
&& {\cal L}\left[\int_0^v p(w)q(v-w)dw;s\right]={\cal L}[p;s]{\cal L}[q;s], \\
\label{Laplace_inverse}
&& {\cal L}^{-1}\left[{\cal L}[p;s]{\cal L}[q;s];v\right] = \int_0^v p(w)q(v-w)dw.
\ea
\end{subequations}

Applying the first form to Eq.\ (\ref{G_tilde_v}), we find that the product of the Laplace transforms of the   factors in the convolution integral is equal to the Laplace transform of $\hat{f}$, or solving for the $G$-dependent factor, that
\be
\label{Laplace_G1}
{\cal L}\left[\hat{\frakg}(v,Q^2);s\right] = \left({\cal L}\left[\hat{h};s\right]\right)^{-1}{\cal L}\left[\hat{f}(v,Q^2);s\right].
\ee
Thus, inverting the Laplace transform on the left, 
\be
\label{Laplace_solution1}
\hat{\frakg}(v,Q^2) = {\cal L}^{-1}\left[ \left({\cal L}\left[\hat{h};s\right]\right)^{-1}{\cal L}\left[\hat{f}(v,Q^2);s\right];v\right].
\ee

The rightmost factor in this equation is not known analytically, so we cannot invert the Laplace transform analytically to determine $\hat{\frakg}(v,Q^2)$.  However, the Laplace transform of $\hat{h}$ is easy to calculate,
\be
\label{Lh}
{\cal L}\left[\hat{h};s\right] = \frac{s^2+3s+4}{(s+1)(s+2)(s+3)}
\ee
as is the (singular) inverse transformation of $\left({\cal L}\left[\hat{h};s\right]\right)^{-1}$,
\be
\label{Lh_inv}
{\cal L}^{-1}\left[\frac{(s+1)(s+2)(s+3)}{s^2+3s+4};v\right] = 3\delta(v)+\delta'(v)-e^{kv}\left(\frac{3}{\omega}\sin \omega v +2\cos\omega v\right).
\ee
Here $k$, $\omega$ are the real and imaginary parts of the roots $\lambda_\pm=k\pm i\omega$ of the polynomial $s^2+3s+4$ in the denominator on the left hand side of Eq.\ (\ref{Lh_inv}).  The $\lambda$'s appeared earlier  as the roots of the differential operator $\left[(\partial/\partial v)^2+3((\partial/\partial v)+4\right]$ in Eq.\ (\ref{v_diff_eq}) as noted in \cite{bdm1}.

We can now use the second form of the convolution theorem in Eq.\ (\ref{Laplace_inverse}) to evaluate the right hand side of Eq.\ (\ref{Laplace_solution1}) as a convolution, and find that
\be
\label{Laplace_solution2}
\hat{\frakg}(v,Q^2) = 3\hat{f}(v,Q^2)+\frac{\partial}{\partial v}\hat{f}(v,Q^2)-\int_0^v\hat{f}(w,Q^2)e^{k(v-w)}\left(\frac{3}{\omega}\sin \omega(v-w)+2\cos\omega(v-w)\right)dw.
\ee
That is, using \eq{fracg}, 
\be
 \hat{G}(v,Q^2) + \frac{2}{3}\frac{1}{\eta_c}\hat{G}(v-\ln\eta_c,Q^2)+ \frac{1}{6}\frac{1}{\eta_b}\hat{G}(v-\ln\eta_b ,Q^2) = \hat{\cal H}_3(v,Q^2),
 \label{Laplace_solution3}
 \ee
 where
 \ba
\hat{\cal H}_3(v,Q^2) &= & \frac{3}{4}\frac{4\pi}{\alpha_s}\left\{3\hat{\cal F}_2^{\gamma p}(v,Q^2)+\frac{\partial}{\partial v}\hat{\cal F}_2^{\gamma p}(v,Q^2) \right. \nonumber \\
 && \left. -\int_0^v \hat{\cal F}_2^{\gamma p}(w,Q^2)e^{k(v-w)}\left(\frac{3}{\omega}\sin \omega(v-w)+2\cos\omega(v-w)\right)dw\right\}. 
 \label{calH3_Laplace}
 \ea
 This result is equivalent to that in Eq.\ (\ref{v_solution}), with the expression on the right equal to $\hat{\cal H}_3(v,Q^2)$. 

 Transforming back to $x$ space, we get
 \ba
 {\cal H}_3(x,Q^2) &=& \frac{3}{4}\frac{4\pi}{\alpha_s}\left\{3{\cal F}_2^{\gamma p}(x,Q^2)-x\frac{\partial}{\partial x}{\cal F}_2^{\gamma p}(x,Q^2) \right. \nonumber \\
 && \left. -\int_x^1 {\cal F}_2^{\gamma p}(z,Q^2)\left(\frac{x}{z}\right)^{3/2}\left[\frac{3}{\omega}\sin \omega\ln\frac{z}{x}+2\cos\omega\ln\frac{z}{x}\right)\frac{dz}{z}\right\}. 
 \label{calH3_Laplace2}
 \ea
The equivalence of this to the expression in Eq.\  (\ref{calH_3}) may be shown  making repeated partial integrations to eliminate as many as possible of the derivatives and double integrals that appear through ${\cal F}_2^{\gamma p}$.
 
 The importance of the Laplace construction is in its avoidance of the intermediate differential equation.  Further, it can also be applied, using the same formalism,  to the DGLAP evolution equation for $G(x,Q^2)$ and other structure functions, as was shown in \cite{bdm2}.


 \subsection{Solution for $G(x,Q^2)$}\label{subsec:G_solution}
 
In the case of four massless quarks considered in Refs. \cite{bdm1,bdm2} and early treatments of parton distributions, $\eta_c=1$, the second term in  Eq.\ (\ref{x_solution}) is just $(2/3) G(x,Q^2)$, the $b$ quark  term is absent, and Eqs.\ (\ref{x_solution}) and (\ref{H3_alt}) give the exact $n_f=4$ solution for $G(x,Q^2)$ in terms of $F_2^{\gamma p}(x,Q^2)$, i.e., 
\be
\label{massless_case}
G_4(x,Q^2)=\frac{3}{5}{\cal H}_3(x,Q^2)={\cal H}_4(x,Q^2)
\ee
The only further calculation necessary is the evaluation of ${\cal H}_3$ using the global fit to $F_2^{\gamma p}(x,Q^2)$---see footnote 
\footnote{For three massless quarks, $G(x,Q^2)\equiv {\cal H}_3(x,Q^2)$, and we could simply denote ${\cal H}_3$  by $G_3(x,Q^2)$, the massless three-quark gluon distribution. To avoid possible confusion of the subscript 3 with the number $n_f>3$ of active quarks in the following discussion, we will not do so, but will continue to use ${\cal H}_3$ to denote the function determined from $F_2^{\gamma p}$ with the normalization in Eqs.\ (\ref{calG_3}) and (\ref{H3_alt}).}.

The situation is more complicated for massive $c$ and $b$ quarks, and we need to solve \eq{Laplace_solution3} (or equivalently, one of  Eqs.\ (\ref{v_solution}) or (\ref{x_solution})) for $G$. We will initially use the equation in the $v$ form, in this case explicitly introducing the threshold step functions, i.e., 
\be
\label{v_solution2}
\hat{G}(v,Q^2)\theta(v)+\frac{2}{3}\frac{1}{\eta_c}\theta(v-\ln \eta_c) \hat{G}(v-\ln\eta_c,Q^2)+\frac{1}{6}\frac{1}{\eta_b}\theta(v-\ln \eta_b)\hat{G}(v-\ln\eta_b,Q^2) = \theta(v) \hat{\cal H}_3(v,Q^2).
\ee
The second and third terms on the left are just translates of the first in $v$-space. The allowable ranges of  the translations are limited by the threshold condition;  the first argument of $\hat{G}$ must be be positive, with $\hat{G}(w,Q^2)\equiv 0$ for  $w\leq  0$. Similarly, $\hat{\cal H}_3(w,Q^2)\equiv 0$ for $w\leq 0$. Having shown these ranges explicitly by introducing the appropriate step functions $\theta(w)$ in the various terms in 
Eq.\ (\ref{v_solution2}), we will now suppress  them in order to keep our expressions simple.

Let $\alpha  = (2/3\eta_c)$, $\beta =(1/6\eta_b)$, and introduce the translation operator $T$ with the property $T(u)f(v)=f(v+u)$. Then Eq.\ (\ref{v_solution2}) can be written as
\be
\label{v_solution3}
\left[1+\alpha T(-\ln\eta_c)+\beta T(-\ln\eta_b)\right]\hat{G}(v,Q^2) = \hat{\cal H}_3(v,Q^2),
\ee
so
\be
\label{v_solution4}
\hat{G}(v,Q^2) = \left[1+\alpha T(-\ln\eta_c)+\beta T(-\ln\eta_b)\right]^{-1}\hat{\cal H}_3(v,Q^2).
\ee
The translations can be implemented explicitly with our Laplace transform technique using the identity
\be
\label{L_translate}
{\cal L}[\theta(v-w)\hat{G}(v-w,Q^2);s]=e^{-ws}{\cal L}[\hat{G}(v,Q^2);s].
\ee
Taking the Laplace transform of Eq.\ (\ref{v_solution2}), we find that the equation can be written in Laplace space as
\be
\label{L_solution1}
\left(1+\frac{2}{3\eta_c}e^{-s\ln\eta_c }+\frac{1}{6\eta_b}e^{-s\ln\eta_b}\right){\cal L}[\hat{G}(v,Q^2);s]={\cal L}[\hat{\cal H}_3(v,Q^2);s],
\ee
so since  we earlier set $\alpha =2/3\eta_c$ and $\beta =1/6\eta_b$,
\be
\label{L_solution2}
{\cal L}[\hat{G}(v,Q^2);s]=\left(1+\alpha e^{-s\ln\eta_c }+\beta e^{-s\ln\eta_b}\right)^{-1} {\cal L}[\hat{\cal H}_3(v,Q^2);s].
\ee

The results in Eqs. (\ref{v_solution4}) and (\ref{L_solution2}) are exact---see footnote 
\footnote{The operator $ \left[1+aT(-\ln\eta_c)+bT(-\ln\eta_b)\right]^{-1}$ on the right-hand side of Eq.\ (\ref{v_solution4}) is the same as that which appeared earlier in Eqs.\ (\ref{c_approx2}) and (\ref{b_approx2}), where $T$ was regarded as a scaling operator in $x$-space.  The equivalence is clear: a translation $v\rightarrow v-\ln \eta$ converts $x=e^{-v}$ to $\eta x$, the scaling transformation used earlier.}.

To make use of these results, we expand the operators on the right in powers of $\alpha$ and $\beta$, and either use the formal properties of $T$ in the case of Eq.\ (\ref{v_solution4}), or calculate the inverse Laplace transform of the  series using the result in Eq.\ (\ref{L_translate}) in the case of Eq.\ (\ref{L_solution2}), to obtain
\begin{subequations}
\ba
\label{v_solution5a}
\hat{G}(v,Q^2) &=&  \hat{\cal H}_3(v,Q^2)+\sum_{n=1}^N (-1)^n[\alpha T(-\ln\eta_c)+\beta T(-\ln\eta_b)]^n\hat{\cal H}_3(v,Q^2)
 \\ 
 \label{v_solution5b}
 &=& \hat{\cal H}_3(v,Q^2)+ {\cal L}^{-1}\left[\sum_{n=1}^N (-1)^n\left(\alpha e^{-s\ln\eta_c }+\beta e^{-s\ln\eta_b}\right)^n{\cal L}\left[\hat{\cal H}_3(v,Q^2);s\right]\right] \\ 
\label{v_solution5c}
&=&  \hat{\cal H}_3(v,Q^2)+\sum_{n=1}^N (-1)^n\sum_{k=0}^n\binom{n}{k}\alpha^{n-k}\beta^k \hat{\cal H}_3(v-(n-k)\ln\eta_c-k\ln\eta_b,Q^2),
\ea
\end{subequations}
where $\binom{n}{k}$ is a binomial coefficient. 

As a consequence of the threshold condition that ${\hat G}(w,Q^2)\equiv 0$ for the argument $w\leq 0$, the summations in Eq.\ (\ref{v_solution5c}) are {\em finite}. Since $\eta_c<\eta_b$ for fixed $Q^2$, the sum on $k$ terminates  for fixed $v$ and $n$ at the smallest $k$ for which $n\ln\eta_c+k(\ln\eta_b-\ln\eta_c)> v$.  Similarly, the sum on $n$ terminates at  $N$ such that $(N+1)\ln\eta_c \geq v$. The result is equivalent to an iterative solution of Eq.\ (\ref{v_solution2}) starting with $\hat{G}_0=\hat{\cal H}_3$.

Converting back to the more familiar variable $x$, we have
\be
\label{x_equation}
G(x,Q^2)+\frac{2}{3}\frac{1}{\eta_c}G(\eta_c x,Q^2)+\frac{1}{6}\frac{1}{\eta_b}G(\eta_b x,Q^2)={\cal H}_3(x,Q^2),
\ee
with the exact solution
\be
\label{x_solution2} 
G(x,Q^2)= H_3(x,Q^2)+\sum_{n=1}^N (-1)^n\sum_{k=0}^n\binom{n}{k}\alpha^{n-k}\beta^k {\cal H}_3(\eta_c^{n-k}\eta_b^kx,Q^2).
\ee
The sums terminate when $x\eta_c^n(\eta_b/\eta_c)^k>1$ ($k$) and $x\eta_c^n>1$ ($n$).

The expression in either Eq.\ (\ref{v_solution5c}) or Eq.\ (\ref{x_solution2}) is exact, but may involve a large number of terms before the series cut off exactly for $Q^2$ large and $x$ small.  For example, for $Q^2=5$ GeV$^2$,  $\eta_c=2.25$, and 14 terms are required for the $c$ series to terminate exactly for $x=10^{-4}$, but only 5 terms for $x=10^{-2}$. For $Q^2=100$ GeV$^2$, $\eta_c=1.0625$, $\ln\eta_c=0.0606$, and the numbers of terms necessary increase to 152 and 75. Fortunately, the factors $\alpha^{n-k}\beta^k$ decrease rapidly, and very good accuracy can be attained in the sums for much smaller values of $N$.

A different approach that is better for much of the region of interest is to rewrite the operator in Eq.\ (\ref{v_solution4}) as
\be
\label{new_iteration}
\frac{1}{1+\alpha T_c+\beta T_b} = \frac{1}{1+\alpha+\beta}\ \frac{1}{1+\alpha'(T_c-1)+\beta'(T_b-1)},
\ee
where $T_c=T(-\ln\eta_c)$, $T_b=T(-\ln\eta_b)$, and
\be
\label{new_parameters}
\alpha'=\alpha/(1+\alpha+\beta ),\qquad \beta'=\beta/(1+\alpha+\beta),
\ee
and then expand the last factor. For $\ln\eta$ small, $T(-\ln\eta)$ approaches the unit operator, and one obtains a rapidly convergent series. Specifically,
\begin{subequations}
\ba
\label{new_iteration2}
\hat{G}(v,Q^2) &=&  \frac{1}{1+\alpha+\beta }\left(\hat{\cal H}_3(v,Q^2)+\sum_{n=1}^N (-1)^n[\alpha'(T_c-1)+\beta'(T_b-1)]^n\hat{\cal H}_3(v,Q^2)\right)\\
&=& \frac{1}{1+\alpha +\beta }\left( \hat{\cal H}_3(v,Q^2)+\sum_{n=1}^N (-1)^n\sum_{k=0}^n\binom{n}{k}[\alpha'(T_c-1)]^{n-k}[\beta'(T_b-1)]^k \hat{\cal H}_3(v,Q^2)\right).
\ea
\end{subequations}
Alternatively,
\be
\label{x_solution3}
G(x,Q^2)=\frac{1}{1+\alpha+\beta }\left({\cal H}_3(x,Q^2)+\sum_{n=1}^N(-1)^n\sum_{k=0}^n\binom{n}{k}[\alpha'(T_c-1)]^{n-k}[\beta'(T_b-1)]^k {\cal H}_3(x,Q^2)\right),
\ee
where $T_i^m{\cal H}_3(x,Q^2)={\cal H}_3(\eta_i^m x,Q^2)$.
The results in these equations are still exact.

It is easily seen that $(T-1)$ acts essentially as a derivative operator around a shifted point:
\be
\label{(T-1)}
\left[T(u)-1\right)]f(v)=\left(T^{1/2}(u)-T^{-1/2}(u)\right)T^{1/2}(u)f(v)=f(v'+ u/2)-f(v'-u/2)\approx uf'(v')
\ee
where $v'=v+u/2$. More generally, $\left[T(u)-1\right]^m f(v)\approx u^m f^{(m)}(v+(m/2)u)$, where $f^{(m)}=d^mf(v)/dv^m$. 

In the present case, the function $\hat{\cal H}_3(v,Q^2)$ on which $T$ operates is well parametrized for $v\agt 0.9 $ ($x\alt 0.1$) by a low-order polynomial in $v$ \cite{bdm1}, and $(T-1)^m\hat{\cal H}_3(v,Q^2)$ decreases rapidly with increasing $m$. The convergence is further enhanced at large $Q^2$ by the smallness of  $\ln\eta_i\approx (4M_i^2/Q^2)$. We have found in practice that expansion through $(T-1)^3$ is generally enough to obtain the accuracy needed. 

We note finally that for $Q^2\gg 4M_b^2\approx 20$ GeV$^2$, $\alpha\approx 2/3$, $\beta \approx 1/6$, and $\hat{\cal H}_3(v,Q^2)/(1+\alpha+\beta )\approx \hat{\cal H}_5(v,Q^2)$, and the expansion starts in the limit of five massless quarks, as  expected.


 \section{Applications}\label{sec:applications}
 
 \subsection{Global parametrization of $F_2^{\gamma p}(x,Q^2)$ using ZEUS structure function data}\label{subsec:ZEUS}
 
As an application of the procedures above, we will numerically  investigate the effects of the $c$ and $b$ masses  on $G(x,Q^2)$ using an updated version of the global parametrization of the ZEUS data for the proton structure function  $F_2^{\gamma p}(x,Q^2)$ \cite{ZEUS1,ZEUS2} made  by Berger, Block and Tan \cite{bbt2}.  Those authors showed that the  data for $x\leq x_P\approx 0.09$ could be parameterized very well as a function of  $Q^2$ and $x$  
with the expression---see footnote 
\footnote{The form of this expression was motivated by the argument that the $x$ dependence of the DIS proton structure functions $F^{\gamma p}_2(x, Q^2)$  should be consistent with saturation of the rigorous Froissart bound \cite{froissart} on hadronic total cross sections, which can increase no more rapidly than $\ln^2(s)\propto \ln^2(1/x)$ for $x\rightarrow 0$. This behavior is in fact observed for a variety of total hadronic cross sections, including $\sigma_{\gamma p}$ \cite{BlockHalzen}, but need not apply to the large-momentum-transfer deep inelastic cross sections until the hard $\gamma^*$-parton cross sections saturate. The model is nevertheless strikingly successful. },
\ba
F_2^{\gamma p}(x,Q^2)=(1-x)&& \hspace*{-1em} \Bigg\{\frac{F_P}{1-x_P}+A(Q^2)\ln\left[\frac {x_P}{x}\frac{1-x}{1-x_P}\right] \nonumber \\
& + & 
B(Q^2)\ln^2\left[\frac {x_P}{x}\frac{1-x}{1-x_P}\right]\Bigg\}. \label{Fofx} 
\ea
Here $x_P$ specifies the location in $x$ of an approximate fixed point observed in the data where curves of $F_2^{\gamma p}(x,Q^2)$ for different $Q^2$ cross. At that point, $\partial F_2^{\gamma p}(x_P,Q^2)/\partial \ln Q^2\approx 0$ for all $Q^2$;  $F_P=F_2^{\gamma p}(x_P,Q^2)$ is the common value of $F_2^{\gamma p}$.  The $Q^2$ dependence of $F_2^{\gamma p}(x,Q^2)$ is  given in those fits by
\ba 
    A(Q^2)&=&a_0+a_1\ln Q^2 +a_2\ln^2 Q^2, \nonumber \\ 
    B(Q^2)&=&b_0+b_1\,\ln Q^2 +b_2\,\ln^2 Q^2.  \label{AB}
\ea

The original fits to DIS data \cite{ZEUS1,ZEUS2} given by Berger, Block, and Tan \cite{bbt2} included data at 24   values of $Q^2$, $Q^2 = 0.11,$ 0.25, 0.65, 2.7, 3.5, 4.5, 6.5, 8.5,  10, 12, 15, 18, 22, 27, 35, 45, 70, 90, 120, 200, 250, 450, 800, and 1200 GeV$^2$, and all $x<x_P$, with the scaling point values $x_P=0.09$ and $F_P=0.41$ assumed to be fixed. The fits were performed  using the ``sieve'' algorithm \cite{sieve}, which minimizes the squared Lorentzian
\be
\Lambda^2_0(\alphabold;\xbold)\equiv\sum_{i=1}^N\ln\left\{1+0.179\delchisq\right\},\label{lambda0}
\ee
where $\chi^2(\alphabold;\xbold)\equiv\sum_{i=1}^N\delchisq $, $\delchisq\equiv 
\left(\left[\bar y_i(x_i;\alphabold)-y_i(x_i)\right]/\sigma_i\right)^2$, $\alphabold$ 
is the parameter space vector, and $\bar y_i(x_i;\alphabold)$ is the theoretical value 
of the measured $y_i$ at $x_i$, with measurement error $\sigma_i$, using a $\delchimax$ cut of 6, to exclude outliers.  

In this paper, we extend the earlier calculation\ \cite{bbt2}, again using the functional forms given in Eqs.\ (\ref{Fofx}) and (\ref{AB}), but modifying the fit  in two important aspects:
First, we fit more ZEUS  data---29 different $Q^2$ values---now covering a larger virtuality range $0.11 \le Q^2\le 2000$ GeV$^2$, using all of the published ZEUS data sets \cite{ZEUS1,ZEUS2} with $x\leq 0.09$, i.e., fitting data for $Q^2=2.7$, 3.5, 4.5, 6.5, 8.5, 10, 12, 15, 18, 22, 27, 35, 45, 60, 70, 90, 120, (150), 200, 250, (350), 450, (650), 800, 1200 , (1500) and (2000) GeV$^2$. The additional data sets that are now included in the fit are given in parentheses. 

Second, we now {\em fit} the scaling point ($x_P,F_P$), leaving the values of $x_P$ and $F_P$ as  free parameters, to be determined by the sieve algorithm.  We  now find the best value for the scaling points to be  
  $x_P =0.0494\pm 0.0039 , F_P =0.503 \pm 0.012$, values significantly different from those assumed in the earlier work \cite{bbt2}.

The new data set has  a total of 210 datum points; using the sieve algorithm \cite{sieve} eliminated 6 points whose total   $\chi^2$ contribution was 61.0. 

\begin{figure}[ht]
\begin{center}
\mbox{\epsfig{file=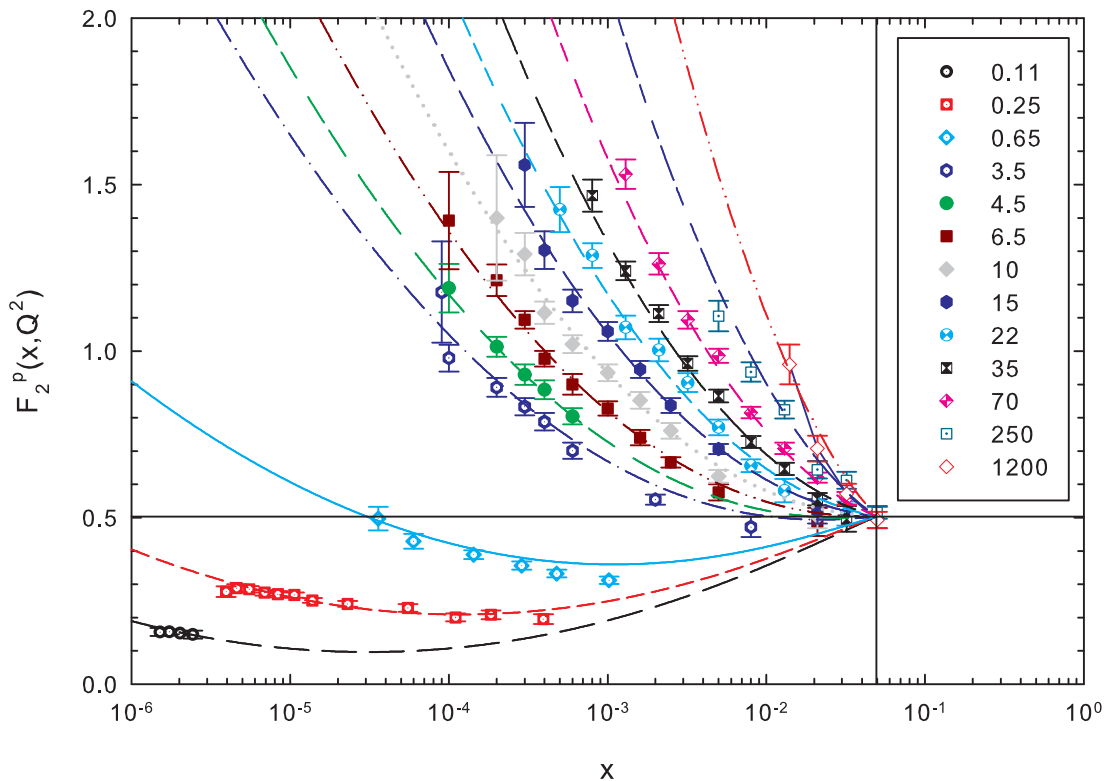,width=6in%
,,bbllx=75pt,bblly=250pt,bburx=510pt,bbury=560pt,clip=%
}}
\end{center}
\caption[]{Plots of the fitted proton structure function, $F_2^{\gamma p}(x,Q^2)$ vs. Bjorken $x$, for virtualities $Q^2=0.11$, 0.25, 0.65, 3.5, 4.5, 6.5, 10, 15, 22, 35, 70, 250 and 1200 GeV$^2$.   The data are from the ZEUS collaboration \cite{ZEUS1,ZEUS2}. The curves are from the fit to the full data sample. The vertical and horizontal lines intersect at the scaling point $x_P=0.049$ and $F_P=0.50$ determined by the fit to the data.}\label{figure:F2p}
\end{figure}

The results of the fit are shown in Fig. \ref{figure:F2p}, where we plot $F_2^{\gamma p}(x,Q^2)$ vs. $x$.  In order not to visually clutter  Fig. \ref{figure:F2p}, only 13  of the 29 sets used in the fit are shown, in the $Q^2$ range from 0.11 to 1200 GeV$^2$, and Bjorken-$x$ range, $10^{-6} \le x\le x_P=0.049$. We see that  the compact fit  of \eq{Fofx} gives a   good  parametrization of the ZEUS experimental data  for the proton structure function $F_2^{\gamma p}(x,Q^2)$ over the available  range of data for small $x$.  It should be stressed that {\em all} curves, independent of their virtuality, go through the scaling point, the intersection of the horizontal and vertical straight lines, providing  a powerful constraint on the fit. The fit satisfies this constraint with  a quite satisfactory goodness-of-fit probability, as seen from Table \ref{table:results}.

\begin{table}[ht]                   
%
\begin{center}
\def\arraystretch{1.2}            
     \caption{\label{fitted}\protect\small Results of a 8-parameter fit to $F_2^{\gamma p}(x,Q^2)$ structure function data \cite{ZEUS1,ZEUS2} using the $x$ 
and $Q^2$ behaviors of \eq{Fofx} and \eq{AB}, with $Q^2$ in GeV$^2$.   The renormalized 
$\chi^2_{\rm min}$ per degree of freedom, taking into account the effects of the 
$\delchimax=6$ cut~\cite{sieve}, is given in the row labeled 
${\cal R}\times\chi^2_{\rm min}$/d.f. The errors in the fitted parameters are 
multiplied by the appropriate $r_{\chi2}$\cite{sieve}.\label{table:results}}
\begin{tabular}[b]{|l||c||}     
	\multicolumn{1}{l}{Parameters}&\multicolumn{1}{c} {Values}\\
\hline
      $a_0$&$-7.828\times 10^{-2}\pm 5.19\times 10^{-3}$ \\ 
      $a_1$&$\phantom{-}2.248\times 10^{-2}\pm 1.47\times 10^{-3}$\\ 
      $a_2$&$\phantom{-}2.301\times 10^{-4}\pm 4.88\times 10^{-4}$\\
\hline
	$b_0$ &$\phantom{-}1.313\times 10^{-2}\pm 6.99\times 10^{-4}$\\
      $b_1$&$\phantom{-}4.736\times 10^{-3}\pm 2.98\times 10^{-4}$\\
      $b_2$&$\phantom{-}1.064\times 10^{-3}\pm 3.88\times 10^{-5}$ \\ 
\hline
$x_P$&$0.0494\pm0.0039$\\
$F_P$&$0.503\pm0.012$\\
	\cline{1-2}
     	\hline
	\hline
	$\chi^2_{\rm min}$&193.19\\
	${\cal R}\times\chi^2_{\rm min}$&215.3\\ 
	d.f.&194\\
\hline
	${\cal R}\times\chi^2_{\rm min}$/d.f.&1.11\\
\hline
\end{tabular}
\end{center}
\end{table}
\def\arraystretch{1}  
 The values of the 8 fit parameters, along with their statistical errors, are given in Table \ref{table:results}. Also shown is the corrected $\chi^2$ per degree of freedom, 1.11,  for 194 degrees of freedom. This yields a  goodness-of-fit probability of 0.15, a reasonable value for this much data. 

Our fit to the data on  $F_2^{\gamma p}(x,Q^2)$ is so far restricted to the region  $x\le x_P$; we have not attempted to fit the (less accurate) DIS data for $x>x_P$. However, the integrals over $F_2^{\gamma p}$ that appear in our expressions for ${\cal H}_3(x,Q^2)$ in Eq.\ (\ref{calH_3}), or in its Laplace equivalent in Eq.\ (\ref{Laplace_solution3}), extend up to $x=1$, so include the interval $x_p<x<1$. In the absence of a global fit to the data in this region, we will simply extend the parametrization, piecewise, to the large-$x$ region, using the form
\ba
F_2^{\gamma p}(x,Q^2)&=&F_P\left(\frac{x}{ x_P}\right)^{\mu(Q^2)}\left(\frac{1-x}{1-x_P}\right)^3,\ \quad x_P<x\le 1, \label{Flarge}
\ea
where the exponent $\mu(Q^2)$ is determined by requiring that the functions in Eqs.\ (\ref{Fofx}) and (\ref{Flarge})  and their  first derivatives with respect to $x$ and $Q^2$ match at $x=x_P$. The results are reasonable, and give the required  parametrization of $F_2^{\gamma p}(x,Q^2)$ over all $x$. The results for $G(x,Q^2)$ in the small-$x$ region $x \ll x_P$ are insensitive  to the contributions to the integrals from the interval $x_P<x\leq 1$, hence,  to the details of the extension of the parametrization.

\subsection{ Evaluation of  ${\cal H}_3(x,Q^2)$.}

We now have a complete parametrization of $F_2^{\gamma p}(x,Q^2)$ using \eq{Fofx} and \eq{AB}, for small $x$, and \eq{Flarge} for large $x$, so we can evaluate ${\calF}_2^{\gamma p}(x,Q^2)$, using Eq.\ (\ref{calF2_1}) or (\ref{calF2}). In these calculations, we use the LO form of $\alpha_s(Q^2)$, 
\be
\alpha_s(Q^2)
=\frac{4 \pi}{\beta_{0}\ln(Q^2/\Lambda^2)},\qquad \beta_{0}= 11-{\frac{2}{3}}n_f,
\label{alphasLO}
\ee
 with $\alpha_s(M_Z^2)=0.118$ and  $\Lambda_5=87.8$ MeV. This is matched to the expression for $n_f=4$ at $Q^2=M_b^2$ with $M_b=4.5$ GeV, giving $\Lambda_4=120.4 $ MeV, which is matched in turn to the expression for $n_f=3$ at $Q^2=M_c^2$ with $M_c=1.3$ GeV, giving $\Lambda_3= 143.6$ MeV.  Finally, we find ${\cal H}_3(x,Q^2)$ by inserting ${ \calF}_2^{\gamma p}(x,Q^2)$ into Eq.\ (\ref{calH_3}) or Eq.\ (\ref{calH3_Laplace2}), and evaluating the integrals numerically. 

We find that we can fit the resulting ${\cal H}_3(x,Q^2)$ very well for small $x$ with an expression quadratic in  both $\ln(Q^2)$ and $\ln (1/ x)$ , with 
\ba
{\cal H}_3(x,Q^2)&=&-2.94-0.359\ln(Q^2)-0.101\ln^2(Q^2)\nonumber\\
&&+\left(0.594-0.0792\ln (Q^2)-0.000578\ln^2(Q^2)   \right)\ln (1/x)\nonumber\\
&&+\left(0.168+0.138\ln (Q^2)+0.0169\ln^2(Q^2)   \right)\ln^2(1/x), \qquad 0<x\le x_G,\label{G3low}
\ea
where $x_G=0.06$.  For the extension to large  $x$, we use 
\ba
{\cal H}_3(x,Q^2)&=&G(x_G,Q^2)\left(\frac{x}{x_G}\right)^{\rho(Q^2)}\left(\frac{1-x}{ 1-x_G}\right)^3,\,\,\qquad x_G<x\le1, \label{G3hi}
\ea
where we determine $\rho(Q^2)$  by matching the first derivatives of \eq{G3low} and \eq{G3hi} at $x=x_G$. 

The function ${\cal H}_3(x,Q^2)$ is determined by the measured $F_2^{\gamma p}(x,Q^2)$, so is taken as fixed when we determine the gluon distribution $G(x,Q^2)$. The transformation to $G(x,Q^2)$ depends on $n_f$ and the quark masses. However, in the simple case of massless quarks, ${\cal H}_3(x,Q^2)$ is just the gluon distribution $G_3(x,Q^2)$ for $n_f=3$ massless quarks. The massless distributions for $n_f=4$ and $n_f=5$ massless quarks are just 3/5 and 6/11 of ${\cal H}_3(x,Q^2)$. We expect the $n_f=3$ and $n_f=5$ massless distributions to represent the  extreme cases of $G(x,Q^2)$ for massive quarks, with the massive gluon distributions approaching $G_3(x,Q^2)$ from below as $Q^2\rightarrow 0$, and $G_5(x,Q^2)$ from above as $Q^2\rightarrow \infty$. This will be evident later  in Fig. \ref{G5ofQsq}. 

We now turn to applications of our results.

\begin{figure}[ht] 
\begin{center}
\mbox{\epsfig{file=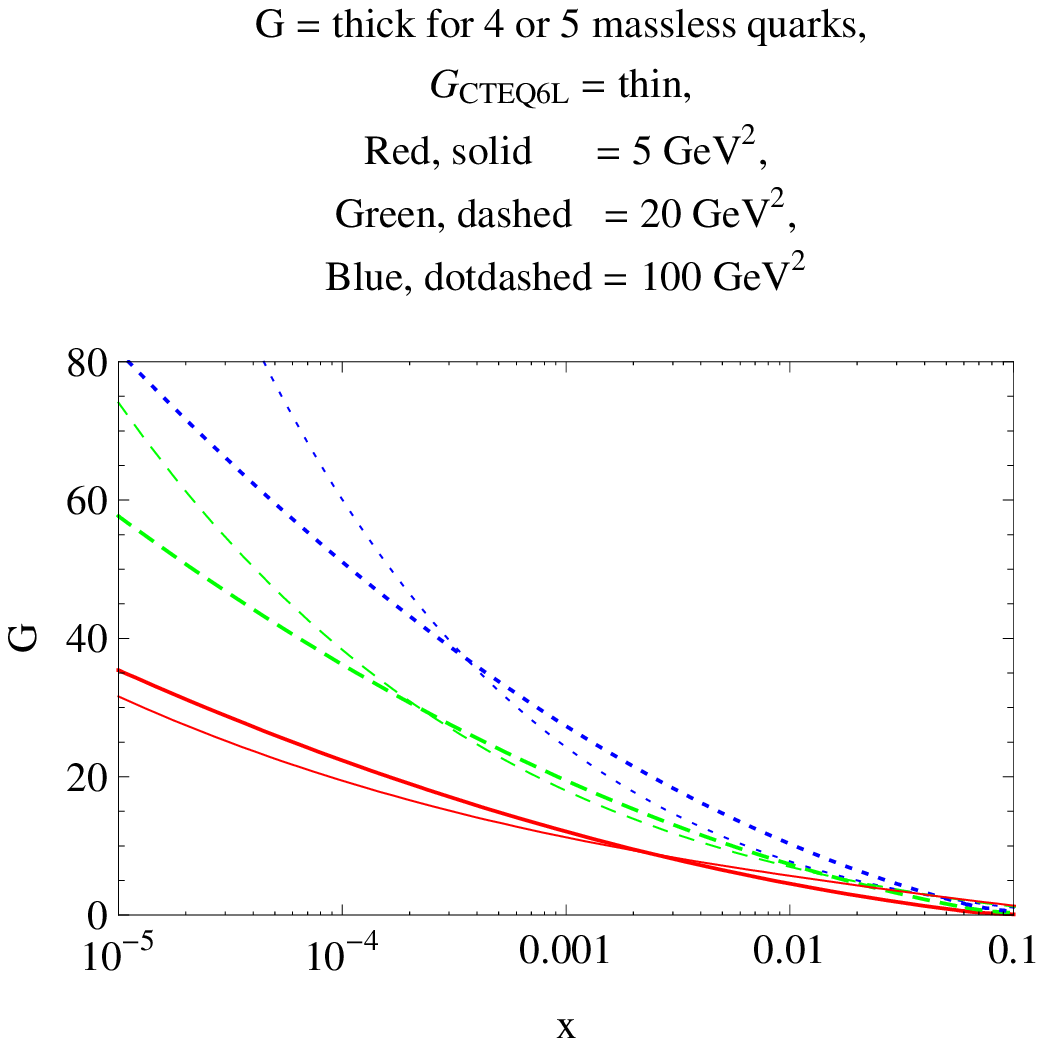
,width=3.5in%
,bbllx=0pt,bblly=0pt,bburx=310pt,bbury=207pt,clip=%
}}
\end{center}
\caption[]{We compare LO $G_{\rm CTEQ6L}(x,Q^2)$ (the thin curves ) with our massless LO $G(x,Q^2)$, for $Q^2= 5$  GeV$^2$ and $n_f=4$ (red solid  curve);  $Q^2= 20 $  GeV$^2$ and $n_f=4$ (green dashed curves); and at $Q^2= 100 $  GeV$^2$ and $n_f=5$ (blue  dotted curves).  The $G_4$ plots for $Q^2=5$ and 20 GeV$^2$ were made using  $G_4=(3/5){\cal H}_3$, and the $G_5$ plot for  $Q^2=100$ GeV$^2$ was made using  $G_5=(6/11){\cal H}_3$, with ${\cal H}_3$ determined from the ZEUS structure function data as described in the text.
}\label{figure:GCTEQ6L}
\end{figure}

\subsubsection{Uncertainties in the LO massless gluon distributions}\label{section:Gerrors}

We immediately deduce from \eq{calH3_Laplace2} that the fractional uncertainty in ${\cal H}_3(x,Q^2)$ is determined by the fractional statistical error in ${\cal F}_2^{\gamma p}(x,Q^2)$, where ${\cal F}_2^{\gamma p}(x,Q^2)$ is given by \eq{calF2_1}. As observed earlier in Section \ref{subsec:heavy_quarks} in our discussion of \eq{F2_evolution2},  the contribution to ${\cal F}_2^{\gamma p}(x,Q^2)$ is dominated by the term 
  $\partial F_2^{\gamma p}(x,Q^2)/\partial\ln Q^2$. Since the LO gluon distribution $G_n(x,Q^2)$ for $n$ massless quarks is a constant multiple of ${\cal H}_3(x,Q^2)$,  we can estimate the fractional statistical error in $G_n$ simply as
\be
\frac{\Delta G_i(x,Q^2)}{ G_n(x,Q^2)}=\frac{\Delta{\cal H}_3(x,Q^2)}{{\cal H}_3(x,Q^2)}\approx\frac {\Delta F_2^{\gamma p}(x,Q^2)}{F_2^{\gamma p}(x,Q^2)},\qquad n=3,\,4,\,5,
\ee
with $\Delta G_i$ and $\Delta F_2^{\gamma p}$  the errors in the functions $G_n$ and the structure function $ F_2^{\gamma p}$, respectively,  with   the errors in the structure function being the statistical errors of the fit parameters. Using standard error analysis techniques, we calculated the fractional statistical  errors $\Delta F_2^{\gamma p}(x,Q^2)/ F_2^{\gamma p}(x,Q^2)$ from the fit parameters  shown in Table \ref{table:results},  using the correlations (not shown) as well as the diagonal elements of the mass squared matrix of the fit.  For $x\la 0.01$,  they were found  to be:  $\la 2\%$ at $Q^2=$ 5 GeV$^2$, $\la 2.5\%$ at $Q^2=$ 20 GeV$^2$, $\la 2.7\%$ for $Q^2=100$ GeV$^2$. We note that for LO massless gluons,  the error estimations are straightforward; the statistical error is effectively the total error for low $x$---no significant approximations are made in this regime    

For comparison, the CTEQ6M total gluon fractional uncertainties \cite{CTEQ6.1} for $Q^2=10$ GeV$^2$ range from  $\sim 20\%$ at  $x=0.0001$ to $\sim 10\%$ at $x=0.01$, errors that are considerably larger. Presumably, this is because of the indirect method that CTEQ employed to get their gluon distributions, solving the coupled DGLAP evolution equations simultaneously for the individual quark and gluon distributiions, and to their inclusion of other data incompatible with the ZEUS data. Their techniques for estimating errors in this more complicated situation are also different.    


\subsubsection{Comparison of the massless $n_f=4$ and $n_f=5$ LO   gluon distributions to  CTEQ6L }\label{subsubsec:CTEQ6_comparison}
The LO CTEQ6L gluon distributions were derived using the complete set of quark and gluon evolution equations, and a scheme for taking mass effects into account for the $c$ and $b$ quarks that depends only on $Q^2$, and does not mix the behavior in $x$ and $Q^2$ as happens in the simplified ACOT scheme. The CTEQ6L treatment of $\alpha_s(Q^2)$ is the same as used here, and as discussed in \cite{bdm2}, the scheme used for treating masses is such that we can  directly  compare  our calculation of $G(x,Q^2)$ for massless quarks using $n_f=4$ to the CTEQ6L result in the region $M_c^2<Q^2\le M_b^2$,  and our  $n_f=5$ massless calculation  to $G_{\rm CTEQ6L}$ for $Q^2>M_b^2$. The results should agree.  We show this comparison in Fig. \ref{figure:GCTEQ6L} for $Q^2=5,\ 20$ and 100 GeV$^2$. The thin curves are those for LO $G_{\rm CTEQ6L}$ \cite{CTEQ6L}, taken from the Durham web site \cite{web}, and the thick curves are for our massless solutions. The solid  red curves are for $Q^2=5$ GeV$^2$, the dashed green curves for $Q^2=20$ GeV$^2$, and the dotted blue curves for $Q^2=100$ GeV$^2$.

These plots of $G$ should be {\em identical} over the region of $x$ and $Q^2$ covered by the experimental ZEUS $F_2^{\gamma p}$ data provided we both fit those data. The minimum $x$-value experimentally observed for a given $Q^2$ are  $x_{\rm min}(5 \ {\rm GeV}^2)=1\times10^{-4},\ x_{\rm min}(20\ {\rm GeV}^2)=5\times10^{-4},$ and $x_{\rm min}(100\ {\rm GeV}^2 )=2\times10^{-3}$. We see from Fig \ref{figure:GCTEQ6L} that, for each of the values of $Q^2$ plotted, there are significant numerical differences for Bjorken-$x$ greater than  $x_{\rm min}(Q^2)$,  i.e., in the regions in which the two determinations of $G(x,Q^2)$ should be the same numerically.  

\begin{figure}[h,t,b] 
\begin{center}
\mbox{\epsfig{file=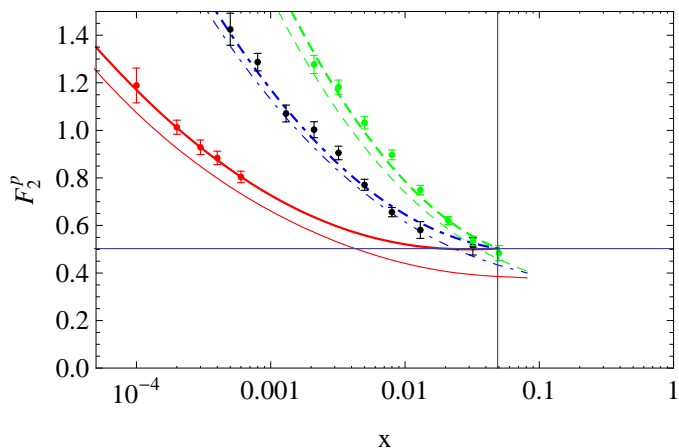
,width=3.5in%
,bbllx=0pt,bblly=0pt,bburx=300pt,bbury=205pt,clip=%
}}
\end{center}
\caption[]{A plot of the proton structure function $F_2^{\gamma p}(x,Q^2)$ vs. $x$, for several virtualities. We  compare our proton structure function fit---the thick curves, using the results of Table \ref{table:results} in \eq{Fofx} and \eq{AB}---to the  $F_2^{\gamma p}(x,Q^2)_{\rm CTEQ6L}$, the thin curves, and  to the ZEUS experimental proton structure function data \cite{ZEUS1,ZEUS2}.  The solid (red) curves are for $Q^2$=4.5 GeV$^2$, the dot-dashed  (blue) curves are for $Q^2$=22 GeV$^2$ and the dashed  (green) curves are for $Q^2$=90 GeV$^2$.  The intersection of the horizontal and vertical lines indicate the scaling point, shown in Table \ref{table:results}. The CTEQ6L values were constructed from their quark distributions \cite{web}.   
}\label{figure:GCTEQ6comp}
\end{figure}

We found earlier \cite{bdm2} that the CTEQ6L gluon distributions $G_{\rm CTEQ6L}(x,Q^2)$ were consistent with the structure functions $F_2^{\gamma p}(x,Q^2)$ {\em calculated} from the CTEQ6L quark distributions using \eq{calH3_Laplace2}. We thus conclude that the CTEQ6L quark distributions must {\em not} give a very  good fit to the ZEUS proton structure function   $F_2^{\gamma p}(x,Q^2)$, which was the starting point of our calculation. This is indeed the case, as is seen in Fig. \ref{figure:GCTEQ6comp}, where thir results systematically miss the experimental data points.  We note that the CTEQ6L fits also use other data, suggesting that incompatibilities in the  distributions required by different data sets may affect their fit to $F_2^{\gamma p}(x,Q^2)$. 

To study this disagreement in more detail, we constructed $F_2^{\gamma p}(x,Q^2)_{\rm CTEQ6L}=\sum_ie_i^2xq_{i,CTEQ6L}(x,Q^2)\ $,
 using the CTEQ6L \cite{CTEQ6L} quark distributions  taken from the Durham web site \cite{web}.
In Fig. \ref{figure:GCTEQ6comp} we show plots   of $F_2^{\gamma p}(x,Q^2)$ vs. $x$ for our fit---using the results of Table \ref{table:results} in \eq{Fofx} and \eq{AB}---as the thick curves and the CTEQ6L fits as the thin curves, comparing them to the   experimental ZEUS proton structure function data \cite{ZEUS1,ZEUS2}. The solid red curves are for  $Q^2=4.5 $ GeV$^2$, the dot-dashed  blue curves are for  $Q^2=22$ GeV$^2$,  and the dashed green curves are for  $Q^2=90 $ GeV$^2$, allowing us to compare virtualities that are very close to those used in Fig. \ref{figure:GCTEQ6L}. Clearly, the CTEQ6L curves are significantly lower than the experimental data, as well as having a different $Q^2$ dependence, resulting in a poor fit to the experimental ZEUS data.  Conversely, our fit is a good representation of the proton  structure function data. This difference explains the discrepancy, in the appropriate $x$ regions,  between the numerical values of the different  gluon distributions shown in  Fig. \ref{figure:GCTEQ6L}. 

We re-emphasize that our method connects $G(x,Q^2)$ in LO {\em directly} to the structure function data, without the need to involve the individual quarks distributions. Other structure function data can be used the same way to derive independent expressions for $G$ \cite{bdm2}. Since the evolution of $G(x,Q^2)$ is built implicitly into the data, we do not need to use the gluon evolution equation to obtain our results. However, we could still use the gluon evolution equation  to check the consistency of our results, e.g., through a determination of $F_{\rm S}$ \cite{bdm2}, giving a test of the adequacy of the LO approximation of the DGLAP equations, a work that is in progress.


\subsubsection{Mass effects in $n_f=4$ distributions}\label{subsubsec:mass_effects}

\begin{figure}[h,t,b] 
\begin{center}
\mbox{\epsfig{file=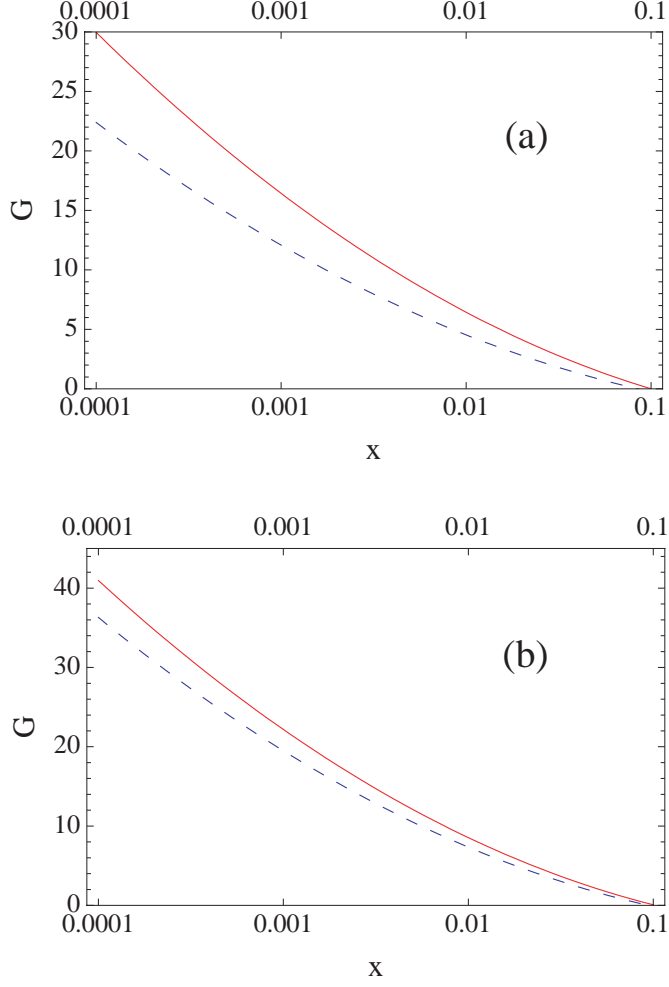
,width=3.5in%
}}
\end{center}
\caption[]{Plots of $G(x,Q^2)$ vs. $x$ for $u,\,d,\,s,$ and $c$ quarks, with the $c$ quark treated as massless (blue dashed curves) and as massive (red solid curves): (a) $Q^2=5$ GeV$^2$; (b) $Q^2=20$ GeV$^2$. The input function ${\cal H}_3(x,Q^2)$ in Eqs.\ (\ref{massless_case}) and (\ref{x_equation}) based on the new global fit to $F_2^{\gamma p}(x,Q^2)$ is given in Eq.\ (\ref{G3low}). The changes in $G$ that result from using the function $F_{2,\rm shifted}^{\gamma p}(x,Q^2)$ with a shifted argument for the $c$ contribution are small, and the resulting curves cannot be distinguished from those shown on the scale of the figure. The differences between the dashed and solid curves result entirely from the mass dependence on the left-hand side of Eq.\   (\ref{x_equation}). }   
\label{figure:GG4comp}
\end{figure}

To investigate the effect of masses on $G(x,Q^2)$ in the simplified ACOT scheme \cite{ACOT}, we have done  model calculations for $Q^2=5$ and 20 GeV$^2$, including only the $c$ quark, and using the approximation of replacing $F_{2,\rm shifted}^{\gamma p}(x,Q^2)$ in Eq.\ (\ref{quark_sum}) by $F_2^{\gamma p}(x,Q^2)$ as in Eq.\ (\ref{F2_evolution2}). The results of the calculations are shown in Fig. \ref{figure:GG4comp}. 

To check the approximation, we have repeated the calculation with $F_2^{\gamma p}$ replaced by $F_2^{\gamma p}(x,Q^2)+\Delta F_2^{\gamma p}(x,Q^2)$, with the shift correction $\Delta F_2^{\gamma p}$ evaluated using the CTEQ6.5 \cite{CTEQ6.5} quark distributions. The results of the modified calculations cannot be distinguished from those in Fig.\ \ref{figure:GG4comp} on the scale of the figure, and are not shown.

The striking mass effects evident in the differences between the solid red (massive) and dashed blue (massless $n_f=4$) curves in the figure result entirely from the mass dependence on the left-hand side of Eq.\ (\ref{x_equation}), with the $b$ term absent.  The convergence of the $(T_c-1)$ expansion for $G(x,Q^2)$ in  Eq.\ (\ref{new_iteration2}) is very rapid, and the magnitude of the difference between the massless and massive curves results largely  from the overall normalization factors, 3/5 and $1/[1+(2/3\eta_c)$ in the equations which relate $G_4={\cal H}_4$ and $G$ to ${\cal H}_3$, Eqs.\ (\ref{massless_case}) and (\ref{x_solution3}).


\subsection{Evaluation of the $n_f=5$ LO gluon distribution $G(x,Q^2)$  for  massive $c$ and $b$ quarks}

 After evaluating the LO $n_f=5$  gluon distribution $G(x,Q^2)$  numerically for massless $u,\,d,\,s$ and  massive $c$ and $b$ quarks using the methods described in Sec.\ \ref{subsec:G_solution}, we found that we could obtain an excellent fit to its $x$ dependence for small $x$ and fixed $Q^2$ using a quadratic expression in $\ln{1/ x}$ or $v$, just as in the case of ${\cal H}_3$ and the related massless distributions. However, the shape of $G(x,Q^2)$ as a function of $Q^2$  is {\em far} from quadratic, and it requires a much more complicated power series in $\ln(Q^2)$ to fit:
\ba
G(x,Q^2)&=&-2.65 - 0.367 \ln(Q^2) + 0.146 \ln^2(Q^2) - 
 0.0321 \ln^3(Q^2) + 
 0.00160\ln^4(Q^2)\nonumber\\
&&+ \left(0.584+ 0.0416 \ln(Q^2) - 0.0666 \ln^2(Q^2) + 
    0.0101 \ln^3(Q^2)- 0.000490 \ln^4(Q^2)\right) \ln(1/ x)\nonumber\\
&&+ \left( 0.155+ 0.123 \ln(Q^2) - 0.0121 \ln^2(Q^2) + 
    0.00270 \ln^3(Q^2)- 0.000111 \ln^4(Q^2) \right) \ln^2(1/x)\nonumber\\
&&\qquad\qquad\qquad\qquad\qquad\qquad\qquad\qquad \qquad\qquad\qquad\qquad\qquad\qquad{\rm for \ } 0<x\le x_G=0.05.\label{finalG5}
\ea

\begin{figure}[h,t,b] 
\begin{center}
\mbox{\epsfig{file=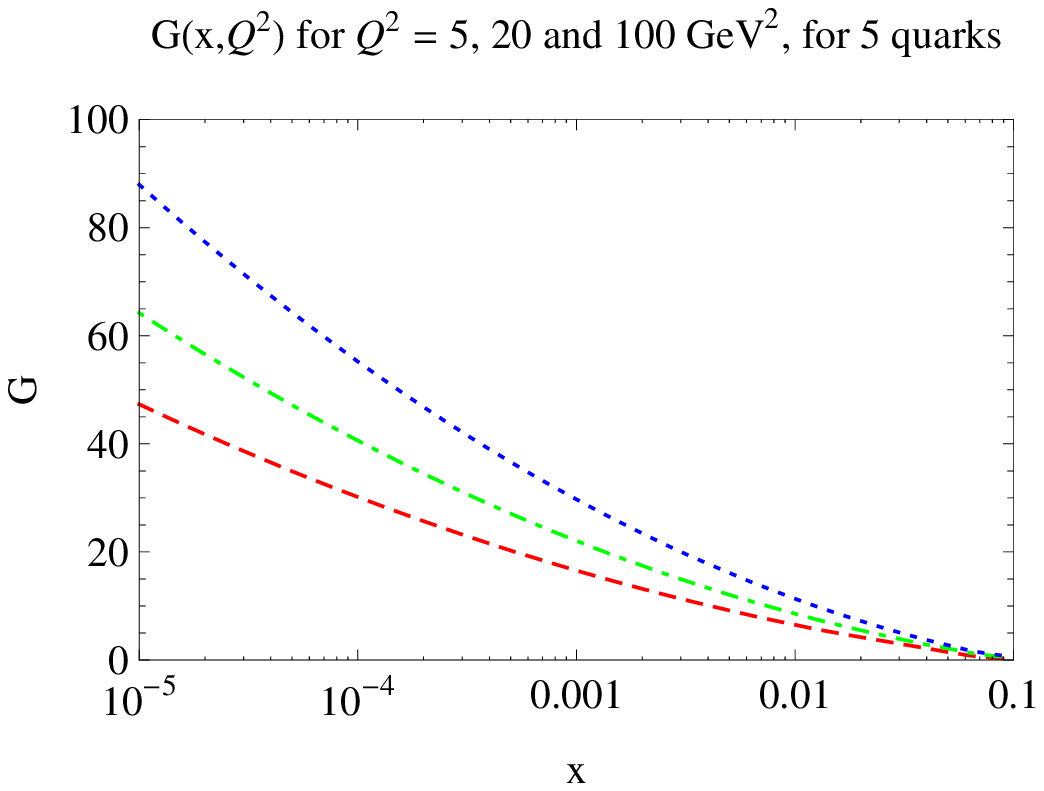
,width=3.5in%
,bbllx=0pt,bblly=0pt,bburx=300pt,bbury=200pt,clip=%
}}
\end{center}
\caption[]{Plots of the final 5-quark gluon distribution for massive quarks, $G(x,Q^2)$ vs. $x$.  We show the LO gluon distributions, $G(x,Q^2)$  for $n_f=5$ with massless $u,d,s$ and massive $c$ and $b$ quarks for: $Q^2= 5$ GeV$^2$ , the dashed red curve; $Q^2= 20$ GeV$^2$, the dot-dashed green curve; and  for $Q^2= 100$ GeV$^2$, the dotted  blue curve. The plots were made from \eq{x_solution2} which used   the value of $ {\cal H}_3(x,Q^2)$ obtained from the new  parametrization of $F_2^P(x,Q^2)$ in Eqs.\ (\ref{Fofx}) and (\ref{AB}) with the parameters in Table \ref{table:results} and the  LO ($n_f=5)$ form of  $\alpha_s(Q^2)$  from Eq.\ (\ref{alphasLO}).}\label{figure:finalG}
\end{figure}

The $x$-dependence  of \eq{x_solution2}, the gluon distribution, $G(x,Q^2)$ for $n_f=5$, with massless $u,d,s$ and massive $c,b$ quarks is plotted  in Fig. \ref{figure:finalG} for  representative values of $Q^2$, namely $Q^2=5,\  20$ and 100 GeV$^2$.  The red dashed  curve was obtained for $Q^2=5$ GeV$^2$, the green dot-dashed  curve for $Q^2=20$ GeV$^2$ and the blue dotted  curve for $Q^2=100$ GeV$^2$.  Again, as was true  for the massless cases, we see that they have a quadratic dependence on $\ln(1/x)$, with the actual value of $G(x,Q^2)$, at {\em any} $x$, lying between the massless cases for $n_f=3$ and $n_f=5$. For $x<0.01$, the systematic error estimates from the approximations that were actually used (see Section \ref{subsec:heavy_quarks}) are comparable with the statistical errors discussed in Section \ref{section:Gerrors}, yielding total errors in the $3-4\%$ range. If we had employed the more accurate approximations discussed in Section \ref{subsec:heavy_quarks},  the errors would have effectively been reduced to the statistical errors of the fit to the ZEUS data \cite{ZEUS1,ZEUS2} for $F_2^{\gamma p}(x,Q^2)$. 

It is evident from the monotonic decrease of $G(x,Q^2)$ with increasing $x$, or, equivalently, the decrease of $\hat G(v,Q^2)$ with decreasing $v$, that the values of  $G(\eta x,Q^2)$ or $\hat{G}(v-\ln\eta)$  at the shifted points that appear in the basic equations, Eqs.\ (\ref{v_solution}) and (\ref{x_solution}) are less than $G(x,Q^2)=\hat{G}(v,Q^2)$.  As a result, using  the fact that $\eta_i>1$,
\be
\label{G_inequal1}
\hat{G}(v,Q^2)+\frac{2}{3}\frac{1}{\eta_c}\hat{G}(v-\ln\eta_c)+\frac{1}{6}\frac{1}{\eta_b}\hat{G}(v-\ln\eta_b) <  (11/6)\hat{G}(v,Q^2).
\ee
The function on the left-hand side of this inequality is equal to the fixed function $\hat{\cal H}_3(v,Q^2)$, so we find that
\be
\label{G_inequal2}
\hat{G}(v,Q^2)>(6/11)\hat{\cal H}_3(v,Q^2)=\hat{G}_5(v,Q^2).
\ee
Also, since 
\be
\label{G_inequal3}
\hat{G}(v,Q^2)+\frac{2}{3}\frac{1}{\eta_c}\hat{G}(v-\ln\eta_c)+\frac{1}{6}\frac{1}{\eta_b}\hat{G}(v-\ln\eta_b)>\hat{G}(v,Q^2),
\ee
we find that $\hat{G}(v,Q^2)<\hat{\cal H}_3(v,Q^2)=\hat{G}_3(v,Q^2)$. 

Thus, for massless $u,\,d,\,s$ quarks and massive $c$ and $b$ quarks, $\hat{G}(v,Q^2)=G(x,Q^2)$, must always lie between the limiting expressions $G_{n_f}(x,Q^2)$ for $n_f=3$ and $n_f=5$ massless quarks. In particular, $G$ approaches $G_5(x,Q^2)$ from above for $Q^2\rightarrow\infty$,  $\ln\eta\sim 4M^2/Q^2\rightarrow 0$, and $G_3(x,Q^2)$ from below for $Q^2\rightarrow 0$, where $\eta x\rightarrow 1$, $G(\eta x,Q^2)\rightarrow0$, and the $b$ and $c$ terms in Eq.\ (\ref{x_equation}) vanish in succession. The limits are, of course, expected: quarks' masses become irrelevant for $Q^2\gg 4M_i^2$, and, conversely,  heavy quarks are not excited at fixed $x$ for $Q^2$ small enough that the threshold condition $x_i=\eta_i x<1$ for pair production from a gluon cannot be satisfied.  

\begin{figure}[ht] 
\begin{center}
\mbox{\epsfig{file=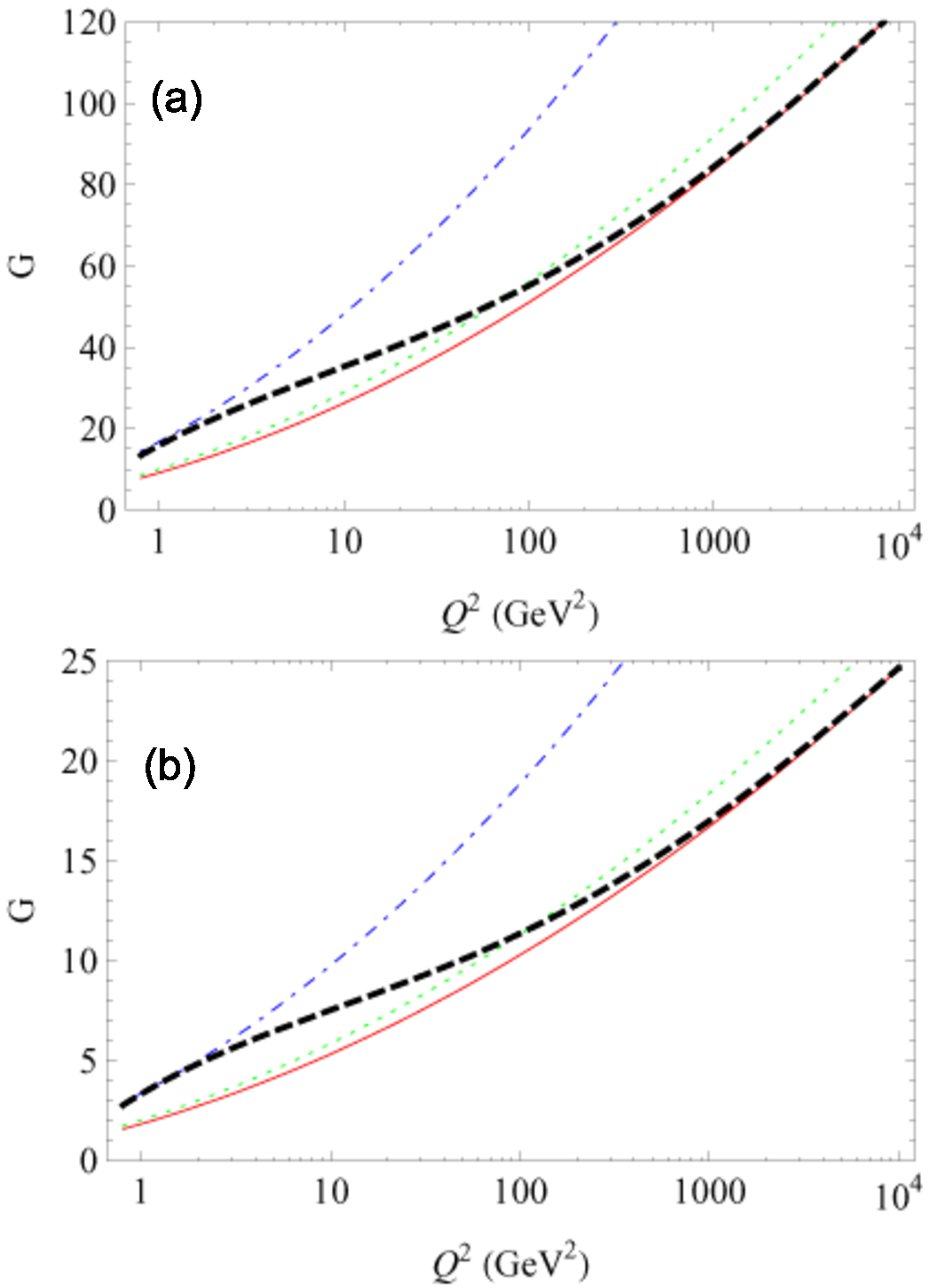
,width=4in%
,bbllx=0pt,bblly=0pt,bburx=300pt,bbury=430pt,clip=%
}}
\end{center}
\caption[]{  $G(x,Q^2)$ vs. $Q^2$, in GeV$^2$, for  (a) $x=10^{-4}$ and (b) $x=10^{-2}$  . The thin dot-dashed blue curve is the $n_f=3$ plot, $G_3$, for massless $u,d,s$ quarks; the thin dotted green curve is the $n_f=4$ plot, $G_4$ for massless $u,d,s,c$ quarks; the thin solid red curve is the $n_f=5$ plot, $G_5$ for massless $u,d,s,c,b$ quarks. The thick  dashed  black curve is $G$, the $n_f=5$ distribution for 3 massless quarks $u,d,s$ and 2 massive quarks $c,b$, from \eq{x_solution2}.  For the evaluation of ${\cal H}_3(x,Q^2)$ that was used in \eq{x_solution2}, we have used the $\alpha_s(Q^2)$ of Eq.\ (\ref{alphasLO}). It should be noted that asymptotically, as $Q^2\rightarrow 0$, $G(x,Q^2)\rightarrow G_3(x,Q^2)$ from below, but as $Q^2\rightarrow \infty$, $G(x,Q^2)\rightarrow G_5(x,Q^2)$ from above, and doesn't appear to significantly overlap  $G_4$ anywhere.
}\label{G5ofQsq}
\end{figure}

This behavior of $G(x,Q^2)$ for massive $c$ and $b$ quarks is shown in Fig.\ \ref{G5ofQsq} over a very wide range of $Q^2$ for two representative values of $x$,  $x=0.0001$ and $x=0.01$.    For comparison, we have also plotted---for the same $x$---the $Q^2$ dependence of the massless $n_f=3$ gluon distribution $G_3(x,Q^2)$ as the thin dot-dashed blue curve, the massless $n_f=4$ gluon distribution $G_4(x,Q^2)$ as the thin dotted green curve, and the massless $n_f=5$ gluon distribution $G_5(x,Q^2)$  as the thin solid red curve. 

 From Figure \ref{G5ofQsq}, we see that all of the massless distributions rise smoothly with $\ln Q^2$ as expected for a dominant $\ln^2Q^2$ behavior as we found for ${\cal H}_3(x,Q^2)$, Eq.\ (\ref{G3low}). However, the thick black dashed curve, the $n_f=5$  distribution $G(x,Q^2)$ for massive $c$ and $b$ quarks is clearly  far from quadratic. As  $Q^2\rightarrow 0$,  $G(x,Q^2)$ approaches the 3 massless quark distribution from {\em below}, differing from $G_3$ by only  a few percent at $Q^2\sim 1$ GeV$^2$.  At $Q^2\sim 50$ GeV$^2$, $G$ crosses the 4-quark case, $G_4$.   For $Q^2\sim 1000$ GeV$^2$, well over the $b$-quark threshold, $G$ approaches the 5-quark massless solution $G_5$ from {\em above}, all as expected from the discussion above.   In between, we have a rather complicated $Q^2$ dependence.   We find these patterns to be true for all small $x$. 
 
 The deviation of the $Q^2$ dependence of $G(x,Q^2)$ from the smooth behavior of ${\cal H}_3(x,Q^2)$ is largely accounted for by the prefactor $1/(1+\alpha+\beta)$ in Eq.\ (\ref{x_solution3}). The series this multiplies converges very rapidly and is dominated by the leading term. Recalling that $\eta_i=1+(4M_i^2/Q^2)$, we see that the prefactor varies from 1 for $Q^2<<4M_c^2$ to 6/11 for $Q^2>>4M_b^2$, and the leading approximation to $G$ has the limits discussed above.  
 
 The term $1/(1+\alpha+\beta )$ in fact  plays the role of a ratio of the sum $\sum_le_l^2=4/3$ for the three massless quarks $u,\,d,\,s$ and their antiquarks used to normalize ${\cal G}_3$ and its integral ${\cal H}_3$, to a sum of squares of  charges for all the quarks,
 \be
 \label{effective_e^2}
 \frac{4}{3}\left(1+\alpha+\beta \right)=\frac{4}{3}+\frac{8}{9}\frac{1}{\eta_c}+\frac{2}{9}\frac{1}{\eta_b}\equiv\sum_ie_{i,\rm eff}^2.
 \ee
The terms $2e_i^2$  appear in this  sum with weights $1/\eta_i$ which we may associate with the degree to which quark $i$ is active, varying from no excitation for $Q^2<<4M_i^2$ to complete excitation for $Q^2>>4M_i^2$.  $G(x,Q^2)$ is then given in leading approximation for massive quarks by the same expression as for massless quarks, $G(x,Q^2)\approx (4/3)(1/\bar{e}_{\rm eff}^2) {\cal H}_3(x,Q^2)$. To the extent to which we may usefully talk about an effective number of active quarks, $n_{f,\rm eff}$, the preceding argument suggests that it should also be defined in terms of the weights $1/\eta_i$, with  $n_{f,\rm eff}=3+(1/\eta_c)+(1/\eta_b)$.


\section{Conclusions}
We have demonstrated that a parametrization of the ZEUS  experimental data \cite{ZEUS1,ZEUS2} on the proton structure function $F_2^{\gamma p}(x,Q^2)$ as a function of $x$ and $Q^2$ in the domain ${\cal D}(x,Q^2)$ (shown in Fig. \ref{figure:F2p}) is all that is needed to obtain an  analytic solution (in the same domain $\cal D$) for the LO gluon distribution $G_{n_f}(x,Q^2)=xg(x,Q^2)$ for $n_f$ massless quarks, since $G_{n_f}$ is a numerical multiple of a function ${\cal H}_3(x,Q^2)$  that  is completely determined by $F_2^{\gamma p}(x,Q^2)$. Comparison with CTEQ6L gluon distributions in the {\em same} domain---where they should agree---show significant inconsistencies; however, these are explained by the fact that the structure function $F_2^{\gamma p}(x,Q^2)$ {\em constructed} from the CTEQ6L quark distributions disagrees markedly with the ZEUS data in the domain $\cal D$.    

The same procedure, again using {\em only} experimental structure function data,  also gives an excellent approximation to $G(x,Q^2)$ when the $c$ and $b$ quarks are properly treated as massive,  using a simplified ACOT approximation \cite{ACOT}. In that case, $G(x,Q^2)$ is the solution of the equation
\be
G(x,Q^2)+\frac{2}{3}\frac{1}{\eta_c} G(x_c,Q^2)+\frac{1}{6}\frac{1}{\eta_b} G(x_b,Q^2) = {\cal H}_3(x,Q^2)
\ee
where $\eta_i=1+(4M_i^2/Q^2)$ and $G(\eta_ix, Q^2)\equiv 0$ for $\eta_ix\geq 1$. This was derived and solved using two different approaches, one based on a differential equation derived from the DGLAP evolution equation for $F_2^{\gamma p}$, and and a second which uses a Laplace transform method to solve that equation directly.

Taking mass effects into account,  the {\em analytic} solution $ G(x,Q^2)$ for 5 quarks--- three massless $u,\ d,\ s$ quarks and  the  massive  $ c$ and $b$ quarks---is given in terms of ${\cal H}_3(x, Q^2)$ by
\ba
 G(x,Q^2)&=&{\cal H}_{3}(x,Q^2)+\sum_{n=1}^N \sum_{m=0}^n(-1)^n\left(\begin{array}{c}\!n\!\\ \!m\!\end{array}\right)
\left(\frac{2}{ 3 \eta_c}
\right)^{n-m}\left(\frac{1}{ 6 \eta_b}
\right)^{m} {\cal H}_{3}\left (x\eta_c^{n-m} \eta_b^m, Q^2  \right),\nonumber\\
&&\quad\qquad\qquad\qquad\qquad\qquad\qquad\qquad\qquad\qquad\qquad x\eta_c^{n-m} \eta_b^m\le 1.\label{Gbexact2inx2}
\ea

As an application of our approach, we have made an accurate parametrization of $F_2^{\gamma p}(x,Q^2)$ using all available low-$x$ ZEUS proton structure function data \cite{ZEUS1,ZEUS2}, and have evaluated ${\cal H}_3(x,Q^2)$ numerically.
We found that ${\cal H}_3(x,Q^2)$ and the massless solutions for $G$  for small $x$ are expressible  numerically as  relatively simple quadratic functions of {\em both} $\ln(1/ x)$ and $\ln(Q^2)$. In the massive quark case, $G(x,Q^2)$ is still quadratic in $\ln{1/ x}$, but is a much more complicated function of $\ln(Q^2)$ which is bounded from above as $Q^2\rightarrow 0$ by the massless 3 quark distribution $G_3(x,Q^2)$, and from below as  $Q^2\rightarrow \infty$ by the massless 5-quark distribution $G_5(x,Q^2)$.   
The same methods can be used to extract $G(x,Q^2)$ in LO from data on other nucleon structure functions in DIS. 

We are currently working on NLO effects on gluon distributions for both massless and massive quarks, as well as checking the consistency of the  DGLAP evolution equations for $F_2^{\gamma p}(x,Q^2)$ and $G(x,Q^2)$ in LO, using proton structure function data.


\begin{acknowledgments}

{\em Acknowledgements:} The authors would like to thank Prof. Douglas W. McKay for his contributions to portions of this work, and also to thank the Aspen Center for Physics for its hospitality during the time parts of this work were done.  

\end{acknowledgments}

\bibliography{gluonsPRD.bib}

\begin{thebibliography}{24}
\expandafter\ifx\csname natexlab\endcsname\relax\def\natexlab#1{#1}\fi
\expandafter\ifx\csname bibnamefont\endcsname\relax
  \def\bibnamefont#1{#1}\fi
\expandafter\ifx\csname bibfnamefont\endcsname\relax
  \def\bibfnamefont#1{#1}\fi
\expandafter\ifx\csname citenamefont\endcsname\relax
  \def\citenamefont#1{#1}\fi
\expandafter\ifx\csname url\endcsname\relax
  \def\url#1{\texttt{#1}}\fi
\expandafter\ifx\csname urlprefix\endcsname\relax\def\urlprefix{URL }\fi
\providecommand{\bibinfo}[2]{#2}
\providecommand{\eprint}[2][]{\url{#2}}

\bibitem[{\citenamefont{Gribov and Lipatov}(1972)}]{dglap1}
\bibinfo{author}{\bibfnamefont{V.~N.} \bibnamefont{Gribov}} \bibnamefont{and}
  \bibinfo{author}{\bibfnamefont{L.~N.} \bibnamefont{Lipatov}},
  \bibinfo{journal}{Sov. J. Nucl. Phys.} \textbf{\bibinfo{volume}{15}},
  \bibinfo{pages}{438} (\bibinfo{year}{1972}).

\bibitem[{\citenamefont{Altarelli and Parisi}(1977)}]{dglap2}
\bibinfo{author}{\bibfnamefont{G.}~\bibnamefont{Altarelli}} \bibnamefont{and}
  \bibinfo{author}{\bibfnamefont{G.}~\bibnamefont{Parisi}},
  \bibinfo{journal}{Nucl. Phys. B} \textbf{\bibinfo{volume}{126}},
  \bibinfo{pages}{298} (\bibinfo{year}{1977}).

\bibitem[{\citenamefont{Dokshitzer}(1977)}]{dglap3}
\bibinfo{author}{\bibfnamefont{Y.~L.} \bibnamefont{Dokshitzer}},
  \bibinfo{journal}{Sov. Phys. JETP} \textbf{\bibinfo{volume}{46}},
  \bibinfo{pages}{641} (\bibinfo{year}{1977}).

\bibitem[{\citenamefont{Pumplin et~al.}(2002)}]{CTEQ6.1}
\bibinfo{author}{\bibfnamefont{J.}~\bibnamefont{Pumplin}} \bibnamefont{et~al.}
  (\bibinfo{collaboration}{CTEQ}), \bibinfo{journal}{J. High Energy Phys.}
  \textbf{\bibinfo{volume}{07}}, \bibinfo{pages}{012} (\bibinfo{year}{2002}),
  \eprint{hep-ph/0201195}.

\bibitem[{\citenamefont{Tung et~al.}(2007)\citenamefont{Tung, Lai, Belyaev,
  Pumplin, Stump, and Yuan}}]{CTEQ6.5}
\bibinfo{author}{\bibfnamefont{W.~K.} \bibnamefont{Tung}},
  \bibinfo{author}{\bibfnamefont{H.~L.} \bibnamefont{Lai}},
  \bibinfo{author}{\bibfnamefont{A.}~\bibnamefont{Belyaev}},
  \bibinfo{author}{\bibfnamefont{J.}~\bibnamefont{Pumplin}},
  \bibinfo{author}{\bibfnamefont{D.}~\bibnamefont{Stump}}, \bibnamefont{and}
  \bibinfo{author}{\bibfnamefont{C.-P.} \bibnamefont{Yuan}},
  \bibinfo{journal}{J. High Energy Phys.} \textbf{\bibinfo{volume}{02}},
  \bibinfo{pages}{053} (\bibinfo{year}{2007}), \eprint{hep-ph/0611254}.

\bibitem[{\citenamefont{Martin et~al.}(2002)\citenamefont{Martin, Roberts,
  Stirling, and Thorne}}]{MRST}
\bibinfo{author}{\bibfnamefont{A.~D.} \bibnamefont{Martin}},
  \bibinfo{author}{\bibfnamefont{R.~G.} \bibnamefont{Roberts}},
  \bibinfo{author}{\bibfnamefont{W.~J.} \bibnamefont{Stirling}},
  \bibnamefont{and} \bibinfo{author}{\bibfnamefont{R.~S.}
  \bibnamefont{Thorne}}, \bibinfo{journal}{Eur. Phys. J. C}
  \textbf{\bibinfo{volume}{23}}, \bibinfo{pages}{73} (\bibinfo{year}{2002}),
  \eprint{hep-ph/0110215}.

\bibitem[{\citenamefont{Martin et~al.}(2004{\natexlab{a}})\citenamefont{Martin,
  Roberts, Stirling, and Thorne}}]{MRST4}
\bibinfo{author}{\bibfnamefont{A.~D.} \bibnamefont{Martin}},
  \bibinfo{author}{\bibfnamefont{R.~G.} \bibnamefont{Roberts}},
  \bibinfo{author}{\bibfnamefont{W.~J.} \bibnamefont{Stirling}},
  \bibnamefont{and} \bibinfo{author}{\bibfnamefont{R.~S.}
  \bibnamefont{Thorne}}, \bibinfo{journal}{Phys. Lett. B}
  \textbf{\bibinfo{volume}{604}}, \bibinfo{pages}{61}
  (\bibinfo{year}{2004}{\natexlab{a}}), \eprint{hep-ph/0410230}.

\bibitem[{\citenamefont{Martin et~al.}(2004{\natexlab{b}})\citenamefont{Martin,
  Roberts, Stirling, and Thorne}}]{MRST3}
\bibinfo{author}{\bibfnamefont{A.~D.} \bibnamefont{Martin}},
  \bibinfo{author}{\bibfnamefont{R.~G.} \bibnamefont{Roberts}},
  \bibinfo{author}{\bibfnamefont{W.~J.} \bibnamefont{Stirling}},
  \bibnamefont{and} \bibinfo{author}{\bibfnamefont{R.~S.}
  \bibnamefont{Thorne}}, \bibinfo{journal}{Eur. Phys. J. C}
  \textbf{\bibinfo{volume}{35}}, \bibinfo{pages}{325}
  (\bibinfo{year}{2004}{\natexlab{b}}), \eprint{hep-ph/0308087}.

\bibitem[{\citenamefont{Stump et~al.}(2002{\natexlab{a}})}]{CTEQchi2}
\bibinfo{author}{\bibfnamefont{D.}~\bibnamefont{Stump}} \bibnamefont{et~al.}
  (\bibinfo{collaboration}{ZEUS}), \bibinfo{journal}{Phys. Rev. D}
  \textbf{\bibinfo{volume}{65}}, \bibinfo{pages}{014012}
  (\bibinfo{year}{2002}{\natexlab{a}}), \eprint{hep-ph/0101051}.

\bibitem[{\citenamefont{Stump et~al.}(2002{\natexlab{b}})}]{CTEQ_Hessian}
\bibinfo{author}{\bibfnamefont{D.}~\bibnamefont{Stump}} \bibnamefont{et~al.}
  (\bibinfo{collaboration}{ZEUS}), \bibinfo{journal}{Phys. Rev. D}
  \textbf{\bibinfo{volume}{65}}, \bibinfo{pages}{014013}
  (\bibinfo{year}{2002}{\natexlab{b}}), \eprint{hep-ph/0101032}.

\bibitem[{\citenamefont{Block et~al.}(2008{\natexlab{a}})\citenamefont{Block,
  Durand, and McKay}}]{bdm1}
\bibinfo{author}{\bibfnamefont{M.~M.} \bibnamefont{Block}},
  \bibinfo{author}{\bibfnamefont{L.}~\bibnamefont{Durand}}, \bibnamefont{and}
  \bibinfo{author}{\bibfnamefont{D.~W.} \bibnamefont{McKay}},
  \bibinfo{journal}{Phys. Rev. D}  (\bibinfo{year}{2008}{\natexlab{a}}),
  \eprint{arXiv:0710.3212 [hep-ph]}.

\bibitem[{\citenamefont{Block et~al.}(2008{\natexlab{b}})\citenamefont{Block,
  Durand, and McKay}}]{bdm2}
\bibinfo{author}{\bibfnamefont{M.~M.} \bibnamefont{Block}},
  \bibinfo{author}{\bibfnamefont{L.}~\bibnamefont{Durand}}, \bibnamefont{and}
  \bibinfo{author}{\bibfnamefont{D.~W.} \bibnamefont{McKay}}
  (\bibinfo{year}{2008}{\natexlab{b}}), \eprint{arXiv:0808.0201 [hep-ph]}.

\bibitem[{\citenamefont{Breitweg et~al.}(2000)}]{ZEUS1}
\bibinfo{author}{\bibfnamefont{J.}~\bibnamefont{Breitweg}} \bibnamefont{et~al.}
  (\bibinfo{collaboration}{ZEUS}), \bibinfo{journal}{Phys. Lett. B}
  \textbf{\bibinfo{volume}{487}}, \bibinfo{pages}{53} (\bibinfo{year}{2000}).

\bibitem[{\citenamefont{Chekanov et~al.}(2001)}]{ZEUS2}
\bibinfo{author}{\bibfnamefont{S.}~\bibnamefont{Chekanov}} \bibnamefont{et~al.}
  (\bibinfo{collaboration}{ZEUS}), \bibinfo{journal}{Eur. Phys. J. C}
  \textbf{\bibinfo{volume}{21}}, \bibinfo{pages}{443} (\bibinfo{year}{2001}).

\bibitem[{\citenamefont{Stump et~al.}(2003)\citenamefont{Stump, Huston,
  Pumplin, Tung, Lai, Kuhlmann, and Owens}}]{CTEQ6L}
\bibinfo{author}{\bibfnamefont{D.}~\bibnamefont{Stump}},
  \bibinfo{author}{\bibfnamefont{J.}~\bibnamefont{Huston}},
  \bibinfo{author}{\bibfnamefont{J.}~\bibnamefont{Pumplin}},
  \bibinfo{author}{\bibfnamefont{W.}~\bibnamefont{Tung}},
  \bibinfo{author}{\bibfnamefont{H.}~\bibnamefont{Lai}},
  \bibinfo{author}{\bibfnamefont{S.}~\bibnamefont{Kuhlmann}}, \bibnamefont{and}
  \bibinfo{author}{\bibfnamefont{J.}~\bibnamefont{Owens}}, \bibinfo{journal}{J.
  High Energy Phys.} \textbf{\bibinfo{volume}{0310}}, \bibinfo{pages}{046}
  (\bibinfo{year}{2003}), \eprint{[hep-ph/0303013]}.

\bibitem[{\citenamefont{Aivazis
  et~al.}(1994{\natexlab{a}})\citenamefont{Aivazis, Collins, Olness, and
  Tung}}]{ACOT}
\bibinfo{author}{\bibfnamefont{M.~A.~G.} \bibnamefont{Aivazis}},
  \bibinfo{author}{\bibfnamefont{J.~C.} \bibnamefont{Collins}},
  \bibinfo{author}{\bibfnamefont{F.~I.} \bibnamefont{Olness}},
  \bibnamefont{and} \bibinfo{author}{\bibfnamefont{W.-K.} \bibnamefont{Tung}},
  \bibinfo{journal}{Phys. Rev. D} \textbf{\bibinfo{volume}{50}},
  \bibinfo{pages}{3102} (\bibinfo{year}{1994}{\natexlab{a}}),
  \eprint{hep-ph/9312319}.

\bibitem[{\citenamefont{Aivazis
  et~al.}(1994{\natexlab{b}})\citenamefont{Aivazis, Olness, and Tung}}]{AOT}
\bibinfo{author}{\bibfnamefont{M.~A.~G.} \bibnamefont{Aivazis}},
  \bibinfo{author}{\bibfnamefont{F.~I.} \bibnamefont{Olness}},
  \bibnamefont{and} \bibinfo{author}{\bibfnamefont{W.-K.} \bibnamefont{Tung}},
  \bibinfo{journal}{Phys. Rev. D} \textbf{\bibinfo{volume}{50}},
  \bibinfo{pages}{3085} (\bibinfo{year}{1994}{\natexlab{b}}),
  \eprint{hep-ph/9312318}.

\bibitem[{\citenamefont{Kr{\"{a}}mer et~al.}(2000)\citenamefont{Kr{\"{a}}mer,
  Olness, and Soper}}]{S-ACOT}
\bibinfo{author}{\bibfnamefont{M.}~\bibnamefont{Kr{\"{a}}mer}},
  \bibinfo{author}{\bibfnamefont{F.~I.} \bibnamefont{Olness}},
  \bibnamefont{and} \bibinfo{author}{\bibfnamefont{D.~E.} \bibnamefont{Soper}},
  \bibinfo{journal}{Phys. Rev. D} \textbf{\bibinfo{volume}{62}},
  \bibinfo{pages}{096007} (\bibinfo{year}{2000}), \eprint{hep-ph/0003035}.

\bibitem[{\citenamefont{Tung et~al.}(2002)\citenamefont{Tung, Kretzer, and
  Schmidt}}]{Tung_masses}
\bibinfo{author}{\bibfnamefont{W.~K.} \bibnamefont{Tung}},
  \bibinfo{author}{\bibfnamefont{S.}~\bibnamefont{Kretzer}}, \bibnamefont{and}
  \bibinfo{author}{\bibfnamefont{C.}~\bibnamefont{Schmidt}},
  \bibinfo{journal}{J. Phys. G} \textbf{\bibinfo{volume}{28}},
  \bibinfo{pages}{983} (\bibinfo{year}{2002}), \eprint{hep-ph/0110247}.

\bibitem[{\citenamefont{Berger et~al.}(2007)\citenamefont{Berger, Block, and
  Tan}}]{bbt2}
\bibinfo{author}{\bibfnamefont{E.~L.} \bibnamefont{Berger}},
  \bibinfo{author}{\bibfnamefont{M.~M.} \bibnamefont{Block}}, \bibnamefont{and}
  \bibinfo{author}{\bibfnamefont{C.-I.} \bibnamefont{Tan}},
  \bibinfo{journal}{Phys. Rev. Lett.} \textbf{\bibinfo{volume}{98}},
  \bibinfo{pages}{242001} (\bibinfo{year}{2007}), \eprint{hep-ph/0703003}.

\bibitem[{\citenamefont{Block}(2006)}]{sieve}
\bibinfo{author}{\bibfnamefont{M.~M.} \bibnamefont{Block}},
  \bibinfo{journal}{Nucl. Inst. and Meth. A.} \textbf{\bibinfo{volume}{556}},
  \bibinfo{pages}{308} (\bibinfo{year}{2006}).

\bibitem[{web()}]{web}
\bibinfo{note}{Http://durpdg.dur.ac.uk/hepdata/mrs.html.}

\bibitem[{\citenamefont{Froissart}(1961)}]{froissart}
\bibinfo{author}{\bibfnamefont{M.}~\bibnamefont{Froissart}},
  \bibinfo{journal}{Phys. Rev.} \textbf{\bibinfo{volume}{123}},
  \bibinfo{pages}{1053} (\bibinfo{year}{1961}).

\bibitem[{\citenamefont{Block and Halzen}(2005)}]{BlockHalzen}
\bibinfo{author}{\bibfnamefont{M.~M.} \bibnamefont{Block}} \bibnamefont{and}
  \bibinfo{author}{\bibfnamefont{F.}~\bibnamefont{Halzen}},
  \bibinfo{journal}{Phys. Rev. D} \textbf{\bibinfo{volume}{72}},
  \bibinfo{pages}{036006} (\bibinfo{year}{2005}).

\end{thebibliography}

\end{document}